\documentclass[a4paper,USenglish]{lipics-v2019}

\usepackage{tikz}
\usetikzlibrary{calc}
  \usetikzlibrary{shadows}
\newcommand{\myurl}[1]{\url{#1}}

\newcommand{\Pre}{{\varphi}}
\newcommand{\Post}{{\psi}}
\newcommand{\Wit}{{\cal{W}}}

\newcommand{\OSC}{{\mathit{OSC}}}
\newcommand{\abs}[1]{| #1 |}
\newcommand{\floor}[1]{\lfloor #1 \rfloor}
\newcommand{\ie}{i.e.,\ }

\newtheorem{mylem}{Lemma}
\newtheorem{mythm}{Theorem}

\newtheorem{mycor}{Corollary}

\usepackage[ruled,noline,noalgohanging]{algorithm2e}
\usepackage{stmaryrd}
\usepackage{isabelle,isabellesym}

\newcounter{manctr}
\newenvironment{manual}{
\setcounter{manctr}{1}
\baselineskip 3.0ex   
\parskip 11pt plus 1pt minus 1pt
\parindent 0pt

}{}

\newcommand{\Ltemplateless}{\texttt{<}}
\newcommand{\Ltemplategreater}{\texttt{>}}

\newcommand{\nspaceunderscore}{\hspace{-0.13em}}

\newcommand{\CC}{C\raise.06ex\hbox{\texttt{++}}}
\newcommand{\CCC}{C\raise.08ex\hbox{\texttt{++}}}
 
\newcommand{\nat}{\hbox{\rm\vrule\kern-0.045em N}}
\newcommand{\real}{\hbox{\rm\vrule\kern-0.035em R}}

\newlength{\setspacing}
\newcommand{\sset}[1]{\{\hspace{\setspacing}#1\hspace{\setspacing} \}}

\newlength{\Labsspacing}
\newcommand{\Labs}[1]{%
\hbox{$| \hspace{\Labsspacing} #1 \hspace{\Labsspacing} |$}}

\newlength{\rowsep} 
\newlength{\entrysep} 

\newlength{\colsep} 
\newlength{\typewidth} 
\newlength{\longtypewidth} 
\newlength{\restoretypewidth} 
\newlength{\callwidth} 
\newlength{\restorecallwidth} 
\newlength{\declwidth} 

\newlength{\longcallwidth}
\newlength{\descriptwidth}
\newlength{\createtextwidth}
\newlength{\textminusdescriptwidth}
\newlength{\typepluscallwidth}

\newcommand{\computewidths}{
          \setlength{\createtextwidth}{\textwidth}
          \addtolength{\createtextwidth}{-\declwidth}             

          \setlength{\longcallwidth}{\textwidth} 
          \addtolength{\longcallwidth}{-\typewidth}

         \setlength{\descriptwidth}{\textwidth}
         \addtolength{\descriptwidth}{-\typewidth}
         \addtolength{\descriptwidth}{-\callwidth} 

         \setlength{\textminusdescriptwidth}{\textwidth}
         \addtolength{\textminusdescriptwidth}{-\descriptwidth}

         \setlength{\typepluscallwidth}{\typewidth}
         \addtolength{\typepluscallwidth}{\callwidth}
       }

\computewidths    

\newlength{\templatestubwidth}
\newlength{\templatebodywidth}
\newlength{\templatelistwidth}

\newcommand{\template}[1]{
\settowidth{\templatestubwidth}{template \Ltemplateless}
\addtolength{\templatestubwidth}{\colsep}
\settowidth{\templatebodywidth}{#1}
\ifthenelse{\templatebodywidth > 0}{%
 \setlength{\templatebodywidth}{\textwidth}
 \addtolength{\templatebodywidth}{-\templatestubwidth}
 \noindent 
 template \Ltemplateless
 \parbox[t]{\templatebodywidth}{#1\Ltemplategreater}
 \settowidth{\templatelistwidth}{#1}
 \ifthenelse{\templatelistwidth > \templatebodywidth}%
 {\vspace{\rowsep}\\}{}
}{}
}

\usepackage{ifthen}
\newlength{\actualcallwidth}
\newlength{\actualtypewidth}
\newlength{\actualtypepluscallwidth}
\newlength{\fnamewidth}
\newlength{\checkwidth}
\newcommand{\returntype}{\,}

\renewcommand{\function}[5][]{ 
 \setlength{\checkwidth}{\typewidth}
 \renewcommand{\returntype}{#2}   
 \addtolength{\checkwidth}{\callwidth}
 \addtolength{\checkwidth}{\descriptwidth}
 \ifthenelse{\lengthtest{\checkwidth = \textwidth}}
            {}
            {\typein{WARNING: The invariant 
\typewidth + \callwidth + \descriptwidth = \textwidth is violated. 
Did you change one of the quantities without calling
\protect\computewidths? I do it for you. If the output looks okay
and you changed textwidth after reading in Lweb.sty it is safe to ignore
this warning. Type any character to proceed.}
\computewidths}
 \settowidth{\actualcallwidth}{#3(#4)}        
 \settowidth{\actualtypewidth}{#2}
 \addtolength{\actualcallwidth}{\colsep}
 \addtolength{\actualtypewidth}{\colsep} 
 \template{#1}
 \ifthenelse{\actualtypewidth > \longtypewidth}
 {
  \parbox[t]{\textwidth}{#2}\\
  \renewcommand{\returntype}{\ }
  \settowidth{\actualtypewidth}{\returntype}
  \addtolength{\actualcallwidth}{\colsep}
 }{}
 \ifthenelse{\actualtypewidth > \typewidth}
 {
  \settowidth{\actualtypepluscallwidth}{\returntype\ #3(#4)}
  \addtolength{\actualtypepluscallwidth}{\colsep}
  \ifthenelse{\actualtypepluscallwidth > \typepluscallwidth}
  {
   \settowidth{\fnamewidth}{\returntype\ #3(}%
   \setlength{\parlistwidth}{\textwidth}%
   \addtolength{\parlistwidth}{-\fnamewidth}%
   \noindent
   \parbox[t]{\textwidth}{%
     \parbox[t]{\fnamewidth}{\returntype\ #3(}%
     \parbox[t]{\parlistwidth}{\raggedright \sloppy #4)}\vspace{\rowsep}\\%
     \hspace*{\typewidth}\hfill\parbox[t]{\descriptwidth}{\sloppy #5 }%
   }
  }
  {
   \parbox[t]{\typepluscallwidth}{\returntype\ #3(#4)}%
   \parbox[t]{\descriptwidth}{\sloppy #5 }%
  }
 }
 {
  \ifthenelse{\actualcallwidth > \callwidth}
  {\settowidth{\fnamewidth}{#3(}%
   \setlength{\parlistwidth}{\longcallwidth}%
   \addtolength{\parlistwidth}{-\fnamewidth}%
   \noindent
   \parbox[t]{\textwidth}{%
    \parbox[t]{\typewidth}{\returntype}\parbox[t]{\fnamewidth}{#3(}%
    \parbox[t]{\parlistwidth}{\raggedright \sloppy #4)}\vspace{\rowsep}\\%
    \hspace*{\typewidth}\hfill\parbox[t]{\descriptwidth}{\sloppy  #5 }%
   }
  }
  {\noindent
   \parbox[t]{\typewidth}{\fussy \returntype}%
   \parbox[t]{\callwidth}{\raggedright \sloppy #3(#4)}%
   \parbox[t]{\descriptwidth}{\sloppy #5 }%
  }
 }
 \vspace{\entrysep}\par
}

\newcommand{\operator}[4][]{%
 \template{#1}
 \settowidth{\actualcallwidth}{#3} 
 \settowidth{\actualtypewidth}{#2}
 \addtolength{\actualcallwidth}{\colsep}
 \addtolength{\actualtypewidth}{\colsep} 
 \ifthenelse{\actualtypewidth > \typewidth}
 {\settowidth{\actualtypepluscallwidth}{#2\ #3}
  \addtolength{\actualtypepluscallwidth}{\colsep}
  \ifthenelse{\actualtypepluscallwidth > \typepluscallwidth}
  {\parbox[t]{\textwidth}{%
    \parbox[t]{\textwidth}{#2\ #3}\vspace{\rowsep}\\%
    \hspace*{\textminusdescriptwidth}\hfill%
    \parbox[t]{\descriptwidth}{\sloppy #4}%
   } 
  }
  {
   \parbox[t]{\typepluscallwidth}{#2\ #3}  
   \parbox[t]{\descriptwidth}{\sloppy #4}%
  }
 }
 {
  \ifthenelse{\actualcallwidth > \callwidth}
  {\noindent
   \parbox[t]{\textwidth}{%
    \parbox[t]{\typewidth}{\fussy #2}
    \parbox[t]{\longcallwidth}{\raggedright\sloppy #3}\vspace{\rowsep}\\%
    \hspace*{\textminusdescriptwidth}\hfill%
    \parbox[t]{\descriptwidth}{\sloppy #4}%
   } 
  }
  {\noindent
   \parbox[t]{\typewidth}{\fussy #2}%
   \parbox[t]{\callwidth}{\raggedright\sloppy #3}%
   \parbox[t]{\descriptwidth}{\sloppy #4}%
  }
 }
 \vspace{\entrysep}\par
}

\usepackage{ifthen}
\newlength{\actualdeclwidth} 
\newlength{\parlistwidth}
\newcommand{\decl}{\,}        
\newcommand{\createpref}{\,}  
\newlength{\actualtypeplusnamewidth}
\newlength{\createtypewidth}

\newcommand{\create}[5][]{
 \setlength{\checkwidth}{\declwidth}   
 \addtolength{\checkwidth}{\createtextwidth}
 \ifthenelse{\lengthtest{\checkwidth = \textwidth}}
 {}
 {\typein{WARNING: The invariant 
\declwidth + \createtextwidth = \textwidth is violated. 
Did you change one of the quantities without calling
\protect\computewidths? I do it for you. If the output looks okay and you
changed textwidth after reading Lweb.sty it is safe to ignore this warning. 
Type any character to proceed.}\computewidths}
 \template{#1}
 \ifthenelse{\equal{#4}{}}%
 {\renewcommand{\decl}{#2\ \ #3;}}%
 {\renewcommand{\decl}{#2\ \ #3(#4);}}
 \settowidth{\createtypewidth}{#2}
 \ifthenelse{\createtypewidth > \longtypewidth}
 {\parbox[t]{\textwidth}{#2}\\
  \ifthenelse{\equal{#4}{}}%
  {\renewcommand{\decl}{\hspace*{\typewidth}#3;}}%
  {\renewcommand{\decl}{\hspace*{\typewidth}#3(#4);}}%
  \renewcommand{\createpref}{\hspace*{\typewidth}#3}
 }{
  \renewcommand{\createpref}{#2\ \ #3}%
 }
 \settowidth{\actualdeclwidth}{\decl}
 \addtolength{\actualdeclwidth}{\colsep}
 \ifthenelse {\actualdeclwidth > \declwidth}
 {
  \ifthenelse{\actualdeclwidth > \textwidth}
  {
   \settowidth{\actualtypeplusnamewidth}{\createpref(}%
   \setlength{\parlistwidth}{\textwidth}%
   \addtolength{\parlistwidth}{-\actualtypeplusnamewidth}%
   \parbox[t]{\textwidth}{%
    \parbox[t]{\actualtypeplusnamewidth}{\createpref(}%
    \parbox[t]{\parlistwidth}{\raggedright #4);}%
    \vspace{\rowsep}\\%
    \hspace*{1cm}\hfill%
    \parbox[t]{\createtextwidth}{\sloppy #5 }%
   }
  }
  {
   \parbox[t]{\textwidth}{%
    \parbox[t]{\textwidth}{\decl}\vspace{\rowsep}\\%
    \hspace*{1cm}\hfill\parbox[t]{\createtextwidth}{\sloppy  #5 }%
   }
  }
 }
 {
  \parbox[t]{\declwidth}{\decl}%
  \parbox[t]{\createtextwidth}{\sloppy #5 }%
 }
 \vspace{\entrysep}\par
}

\usepackage{ifthen}
\newlength{\actualdestructwidth}

\newcommand{\destruct}[2]{
  \settowidth{\actualdestructwidth}{$\sim$#1()}
  \ifthenelse {\actualdestructwidth > \declwidth}
  {\noindent
   \parbox[t]{\textwidth}{%
    \parbox[t]{\textwidth}{$\sim$#1()}\vspace{\rowsep}\\%
    \hspace*{1cm}\hfill\parbox[t]{\createtextwidth}{\sloppy #2}%
   } 
  }
  {\noindent\parbox[t]{\declwidth}{$\sim$#1()}%
   \parbox[t]{\createtextwidth}{\sloppy #2}%
  }
  \vspace{\entrysep}\par 
}

\newlength{\firstcolwidth}

\newcommand{\enum}[3]{%
\settowidth{\firstcolwidth}{#1\ \{\ #2\ \}}
\ifthenelse {\firstcolwidth > \declwidth}{
\noindent\parbox[t]{\textwidth}{%
\parbox[t]{\textwidth}{#1\ \{\ #2\ \}}\vspace{\rowsep}\\%
\hspace*{1cm}\hfill\parbox[t]{\createtextwidth}{\sloppy #3}%
}}
{\noindent\parbox[t]{\declwidth}{#1\ \{\ #2\ \}}%
\parbox[t]{\createtextwidth}{\sloppy #3}%
}\par}

\newcommand{\typemember}[3][]{%
 \settowidth{\firstcolwidth}{#2}%
 \addtolength{\firstcolwidth}{\colsep}%
 \template{#1}
 \ifthenelse {\firstcolwidth > \declwidth}%
 {\noindent%
  \parbox[t]{\textwidth}{%
   \parbox[t]{\textwidth}{#2}\vspace{\rowsep}\\%
   \hspace*{1cm}\hfill\parbox[t]{\createtextwidth}{\sloppy #3}%
  }
 }
 {\noindent\parbox[t]{\declwidth}{#2}%
  \parbox[t]{\createtextwidth}{\sloppy #3}%
 }
 \par%
}

\newcommand{\typedef}[3]{%
\settowidth{\firstcolwidth}{typedef #1 #2}
\ifthenelse {\firstcolwidth > \declwidth}{
\noindent\parbox[t]{\textwidth}{%
\parbox[t]{\textwidth}{typedef #1 #2}\vspace{\rowsep}\\%
\hspace*{1cm}\hfill\parbox[t]{\createtextwidth}{\sloppy #3}%
}}
{\noindent\parbox[t]{\declwidth}{typedef #1\ #2\ }%
\parbox[t]{\createtextwidth}{\sloppy #3}%
}\par}

\newcommand{\event}[2]{%
\settowidth{\firstcolwidth}{#1}
\addtolength{\firstcolwidth}{\colsep}
\ifthenelse {\firstcolwidth > \declwidth}{
\noindent\parbox[t]{\textwidth}{%
\parbox[t]{\textwidth}{#1}\vspace{\rowsep}\\%
\hspace*{1cm}\hfill\parbox[t]{\createtextwidth}{\sloppy #2}%
}}
{\noindent\parbox[t]{\declwidth}{#1}%
\parbox[t]{\createtextwidth}{\sloppy #2}%
}\par}

\newcounter{classwidth}
\newcounter{arrowtip}
\newcounter{arrowline}
\newcounter{indentcol}
\newcounter{framerow}
\newcounter{arrowrow}

\newsavebox{\genarrowbox}
\savebox{\genarrowbox}{%
  \begin{picture}(0,0)%
  \put(0,0){$\triangleleft$}
  \put(1,0.55){\line(1,0){2}}
  \put(3,0.55){\line(0,-1){1}}
  \end{picture}
}

\newlength{\semwidth}
\newlength{\semindent}
\title{Trustworthy Graph Algorithms}
\author{Mohammad Abdulaziz}{TU M\"unchen}{}{}{}
\author{Kurt Mehlhorn}{MPI for Informatics}{}{}{}
\author{Tobias Nipkow}{TU M\"unchen}{}{}{}
\authorrunning{M.~Abdulaziz, K.~Mehlhorn, T.~Nipkow}
\Copyright{Mohammad Abdulaziz, Kurt Mehlhorn, Tobias Nipkow}

\nolinenumbers

\begin{document}

\maketitle

\begin{abstract}
The goal of the LEDA project was to build an easy-to-use and extendable library of correct and  efficient data structures, graph algorithms and geometric algorithms. We report on the use of formal program verification to achieve an even higher level of trustworthiness. Specifically, we report on an ongoing and largely finished verification of the blossom-shrinking algorithm for maximum cardinality matching.
\end{abstract}

\section{Introduction}

This talk is a follow-up on two previous invited MFCS-talks given by the second author:
\begin{itemize}
\item \emph{LEDA: A Library of Efficient Data Types and Algorithms} in MFCS 1989~\cite{Mehlhorn-Naeher:LEDA}, and
\item \emph{From Algorithms to Working Programs: On the Use of Program Checking in LEDA} in MFCS 1998~\cite{Mehlhorn-Naeher:MFCS98}.
\end{itemize}
After a review of these papers, we discuss the further steps taken to reach even higher trustworthiness of our implementations.
\begin{itemize}
  \item Formal correctness proofs of checker programs~\cite{FrameworkVerificationCertifyingComputations,Verification-CertComps-AutoCorres-Simpl}, and
  \item Formal verification of complex graph algorithms~\cite{BlossomCorrectnessProof}.
  \end{itemize}
The second item is the technical core of the paper: it reports on the ongoing and largely finished verification of the blossom-shrinking algorithm for maximum cardinality matching in Isa\-belle/HOL by the first author.

  \subparagraph{Personal Note by the Second Author:} As this paper spans 30 years of work, the reader might get the impression that I followed a plan. This is not the case. As a science, in this case computer science, progresses, there are logical next steps. I took these steps. I did not know 30 years ago, where the journey would lead me.

\section{Level One of Trustworthiness: The LEDA Library of Efficient Data Types and Algorithms} In 1989, Stefan N\"aher and the second author set out to build an easy-to-use and extendable library of correct and  efficient data structures, graph algorithms and geometric algorithms. The project was announced in an invited talk at MFCS 1989~\cite{Mehlhorn-Naeher:LEDA} and the library is available from Algorithmic Solutions GmbH~\cite{LEDAsystem}. LEDA, the library of efficient data types and algorithms, offers a flexible data type graph with loops for iterating over edges and nodes and arrays indexed by nodes and edges. It also offers the data types required for graph algorithms such as queues, stacks, and priority queues. It thus created a framework in which graph algorithms can be formulated easily and naturally, see Figure~\ref{Dijkstra} for an example. The design goal was to create a system in which the difference between the pseudo-code used to explain an algorithm and what constitutes an executable program is as small as possible. The expectation was that this would ease the burden of the implementer and make it easier to get implementations correct.

\begin{figure}[t]
  \small
  \begin{verbatim}
template <class NT>
void DIJKSTRA_T(const graph& G, node s, const edge_array<NT>& cost,
                   node_array<NT>& dist, node_array<edge>& pred)
{
  node_pq<NT>  PQ(G);   // a priority queue for the nodes of G
  node v; edge e; 
  dist[s] = 0;          // distance from s to s is zero
  PQ.insert(s,0);       // insert s with value 0 into PQ
  forall_nodes(v,G) pred[v] = nil; // no incoming tree edge yet
  while (!PQ.empty())   // as long as PQ is non-empty
  { node u = PQ.del_min();   // let u be the node with minimum dist in PQ
    NT du = dist[u];         // and du its distance
    forall_adj_edges(e,u)    // iterate over all edges e out of u                       
    { v = G.opposite(u,e);   // makes it work for ugraphs
      NT c = du + cost[e];   // distance to v via u
      if (pred[v] == nil && v != s )            // v already reached?
        PQ.insert(v,c);                         // first path to v
      else if (c < dist[v]) PQ.decrease_p(v,c); // better path
           else continue;
      dist[v] = c;           // store distance value
      pred[v] = e;           // and incoming tree edge  
    }
   }
  }
\end{verbatim}

  \caption{\label{Dijkstra}The LEDA implementation of Dijkstra's algorithm: Note that the executable code above is similar to a typical pseudo-code presentation of the algorithm. }
\end{figure}

\section{Level Two of Trustworthiness: Certifying Algorithms}
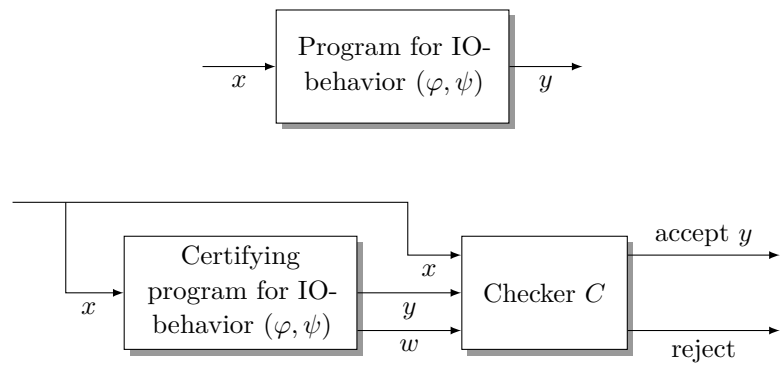
\begin{figure}[t]
\centering
\begin{tikzpicture}
\tikzstyle{box}=[draw, fill=white, drop shadow={opacity=0.8},
  inner xsep=8pt, minimum height=1.5cm];
\node[box,text width=2.5cm,text centered] (P) at (0,5) {Program for IO-behavior $(\Pre,\Post)$};
\draw[-latex] (-2.5,5) -- node[below] {$x$} (P);
\draw[-latex] (P) -- node[below] {$y$} (2.5,5);

\node[box, text width=2.5cm, text centered] (cP) at (-2,2)
  {Certifying program for IO-behavior $(\Pre,\Post)$};
\node[box] (C) at (2,2) {Checker $C$};
\draw[-latex]
  (-5,3.2) -- (0.2,3.2) |- node[below, pos=0.7] {$x$}
  ($(C.west) + (0,0.5)$);
\draw[-latex] (-4.3,3.2) |- node[below, pos=0.7] {$x$} (cP.west);
\draw[-latex] (cP.east) -- node[below] {$y$} (C.west);
\draw[-latex]
  ($(cP.east) - (0,0.5)$) -- node[below] {$w$}
  ($(C.west) - (0,0.5)$);
\draw[-latex] ($(C.east) + (0,0.5)$) -- node[above] {accept $y$} +(2,0);
\draw[-latex] ($(C.east) - (0,0.5)$) -- node[below] {reject} +(2,0);
\end{tikzpicture}
\caption{The top figure shows the I/O behavior of a conventional program
  for IO-behavior $(\Pre,\Post)$; here $\Pre$ is the precondition and $\Post$ is the postcondition. The user feeds an input $x$ satisfying
  $\Pre(x)$ to the program and the program returns an output $y$ satisfying
  $\Post(x,y)$. A certifying algorithm for IO-behavior $(\Pre,\Post)$
  computes $y$ and a witness $w$. The checker $C$ accepts the triple
  $(x,y,w)$ if and only if $w$ is a valid witness for the postcondition
  $\Post(x,y)$, i.e., it proves $\Post(x,y)$. (reprinted from~\cite{FrameworkVerificationCertifyingComputations})}
  \label{certifying alg} 
\end{figure}

Nevertheless, some implementations in the initial releases were incorrect, in particular, the planarity test\footnote{Most of the implementations of the geometric algorithms were also incorrect in their first release as we had na\"\i vely used floating point arithmetic to implement real arithmetic and the rounding errors invalidated the implementations of the geometric primitives.  This lead to the development of the exact computation paradigm for geometric computing by us and others~\cite{ClassRoomExample,yap:crc:03,Fortune96,Yap97,Mehlhorn-Naeher4}. In this paper, we restrict to graph algorithms.}; it declared some planar graphs non-planar. At around 1995, we adopted the concept of certifying algorithms~\cite{Mehlhorn-Naeher:MFCS98,McConnell2010} for the library and reimplemented all algorithms~\cite{LEDAbook}. A certifying algorithm computes for each input a easy-to-check certificate (witness) that demonstrates to the user that the output of the program for this particular input is correct; see Figure~\ref{certifying alg}. For example, the certifying planarity test returns a Kuratowski subgraph if it declares the input graph non-planar and a (combinatorial) planar embedding if it declares the input graph planar, and the maximum cardinality matching algorithm computes a matching and an odd-set-cover that proves its optimality; see Figures~\ref{Matching} and~\ref{Checker}. The state of the art of certifying algorithms is described in~\cite{McConnell2010}. We also implemented checker programs that check the witness for correctness and argued that the checker programs are so simple that their correctness is evident. From a pragmatic point of view, the goals of the project were reached by 2010. The library was easy-to-use and extendable, the implementations were efficient, and no error was discovered in any of the graph algorithms for several years despite intensive use by a commercial and academic user community.

Note that, most likely, errors would not have gone undiscovered because of the use of certifying algorithms and checker programs. Only if a module produced an incorrect output and hence an invalid certificate and the checker program missed to uncover the invalidity of the certificate would an error go unnoticed. Of course, the possibility is  there and the phrase ``most likely'' in the preceding sentence has no mathematical meaning.

Alternative libraries such as Boost and LEMON~\cite{Boost,LEMON} are available now and some of their implementations are slightly more efficient than ours. However, none of the new libraries pays the same attention to correctness. For example, all libraries allow floating point numbers as weights and capacities in network algorithms, but only LEDA ensures that the intricacies of floating point arithmetic do not invalidate the implementations; see~\cite{Althaus-Mehlhorn} and~\cite[Section 7.2]{LEDAbook}.

\begin{figure}[t]
\hrule
  {\small
\begin{manual}\setlength{\parskip}{0.4\parskip}
A \emph{matching} in a graph $G$ is a subset $M$ of the 
edges of $G$ such that no two share an endpoint. 

An odd-set cover \mbox{$\mathit{OSC}$} of $G$ is a labeling of the nodes of $G$ with
non-negative integers such that every edge of $G$ (which is not a self-loop) 
is either incident to a node labeled $1$ or 
connects two nodes labeled with the same $i$, $i \ge 2$. 

Let $n_i$ be the number of nodes labeled $i$ and consider any matching
$N$. For $i$, $i \ge 2$, let $N_i$ be the edges in $N$ that connect 
two nodes labeled $i$.
Let $N_1$ be the remaining edges in $N$. Then 
$\Labs{N_i} \le \lfloor n_i/2 \rfloor$ and $\Labs{N_1} \le n_1$
and hence
\[ \Labs{N} \le n_1 + \sum_{i \ge 2} \lfloor n_i/2 \rfloor \]
for any matching $N$ and any odd-set cover \mbox{$\mathit{OSC}$}.
It can be shown that for a maximum cardinality matching $M$
there is always an odd-set cover \mbox{$\mathit{OSC}$} with 
\[ \Labs{M} = n_1 + \sum_{i \ge 2} \lfloor n_i/2 \rfloor, \]
thus proving the optimality of $M$. In such a cover all $n_i$ with $i \ge 2$
are odd, hence the name.

\settowidth{\typewidth}{\mbox{$\mathit{list}\Ltemplateless \mathit{edge}\Ltemplategreater $}}
\addtolength{\typewidth}{\colsep}
\settowidth{\callwidth}{MAX}
\computewidths

\function {\mbox{$\mathit{list}\Ltemplateless \mathit{edge}\Ltemplategreater $}}
{MAX\nspaceunderscore\_CARD\nspaceunderscore\_MATCHING} 
{\mbox{$\mathit{graph}\ G,$}
\mbox{$\mathit{node\nspaceunderscore\_array}\Ltemplateless \mathit{int}\Ltemplategreater \&\ \mathit{OSC}$}}
{computes a maximum cardinality matching $M$ in $G$ and
returns it as a list of edges.
The algorithm (\cite{Edmonds:matching}, \cite{Gabow:edmonds}) has running 
time $O(nm\cdot\alpha(n,m))$.

An odd-set cover that proves the maximality of $M$ is returned in \mbox{$\mathit{OSC}$}.
\bigskip
 
}

\function {\mbox{$\mathit{bool}$}}
{CHECK\nspaceunderscore\_MAX\nspaceunderscore\_CARD\nspaceunderscore\_MATCHING} 
{\mbox{$\mathit{graph}\ G,$}
\mbox{$\mathit{list}\Ltemplateless \mathit{edge}\Ltemplategreater \ M,$}
\mbox{$\mathit{node\nspaceunderscore\_array}\Ltemplateless \mathit{int}\Ltemplategreater \ \mathit{OSC}$}} 
{checks whether \mbox{$M$} is a maximum cardinality matching in $G$ and
\mbox{$\mathit{OSC}$} is a proof of optimality. Aborts if this is not the case. 
}
\end{manual}\smallskip

\hrule
}\smallskip

\caption{The LEDA manual page for maximum cardinality matchings (reprinted from~\cite{Mehlhorn-Naeher:MFCS98}).}
\label{Matching}
\end{figure}

\begin{figure}[t]
{\small
\begin{verbatim}
static bool return_false(string s)
{ cerr << "CHECK_MAX_CARD_MATCHING: " << s << "\n"; return false; }

bool CHECK_MAX_CARD_MATCHING(const graph& G, const list<edge>& M,
                                    const node_array<int>& OSC)
{ int n = Max(2,G.number_of_nodes());
  int K = 1;
  array<int> count(n);
  for (int i = 0; i < n; i++) count[i] = 0;
  node v; edge e;

  forall_nodes(v,G) 
  { if ( OSC[v] < 0 || OSC[v] >= n ) 
     return_false("negative label or label larger than n - 1");
    count[OSC[v]]++;
    if (OSC[v] > K) K = OSC[v];
  }

  int S = count[1];
  for (int i = 2; i <= K; i++) S += count[i]/2;
  if ( S != M.length() )
    return_false("OSC does not prove optimality");

  forall_edges(e,G)
  { node v = G.source(e); node w = G.target(e);
    if ( v == w || OSC[v] == 1 || OSC[w] == 1 ||
            ( OSC[v] == OSC[w] && OSC[v] >= 2) ) continue;
    return_false("OSC is not a cover");
  }
  return true;
}
\end{verbatim}
}
\caption{The checker for maximum cardinality matchings (reprinted from~\cite{Mehlhorn-Naeher:MFCS98}).}
\label{Checker}
\end{figure}

\section{Level Three of Trustworthiness: Formal Verification of Checkers}
\label{sec:level3}
We stated above that the checker programs are so simple that their correctness is evident. Shouldn't they then be amenable to formal verification? Harald Ganzinger and the second author attempted to do so at around 2000 and failed.  About 10 years later (2011 -- 2014) Eyad Alkassar from the Verisoft Project~\cite{Verisoft}, Sascha B\"ohme and Lars Noschinski from Tobias Nipkow's group at TU M\"unchen, and Christine Rizkallah and the second author succeeded in formally verifying some of the checker programs~\cite{FrameworkVerificationCertifyingComputations,Verification-CertComps-AutoCorres-Simpl}.  In order to be able to talk about formal verification of checker programs, we need to take a more formal look at certifying algorithms. 

We consider algorithms which take an input from a set $X$ and produce an
output in a set $Y$ and a witness in a set $W$. The input $x \in X$ is
supposed to satisfy a precondition $\Pre(x)$, and the input together with
the output $y \in Y$ is supposed to satisfy a postcondition $\Post(x,y)$. 
A \emph{witness predicate} for a specification with precondition~$\Pre$ and
postcondition $\Post$ is a predicate $\Wit \subseteq X \times Y \times W$,
where $W$ is a set of witnesses
with the following \emph{witness property}:
\begin{equation}
\Pre(x) \land \Wit(x,y,w) \longrightarrow \Post(x,y).
\label{witness-property}
\end{equation}
\noindent The checker program $C$ receives a triple\footnote{We ignore the minor complication that $X$, $Y$, and $W$ are abstract sets and programs handle concrete representations.} $(x,y,w)$ and is supposed to
check whether
it fulfills the witness property. 
If $\neg \Pre(x)$, $C$ may do anything (run forever or halt with an
arbitrary output). If $\Pre(x)$, $C$ must halt and either accept or reject.
It is required to accept if $\Wit(x,y,w)$
holds and is required to reject otherwise. This results in the following proof obligations.

\begin{description}
\item[Checker Correctness:]
  We need to prove that $C$ checks the witness predicate 
  assuming that the precondition holds, i.e., on input $(x,y,w)$:
  \begin{enumerate} 
  \item If $\Pre(x)$, $C$ halts.
  \item If $\Pre(x)$ and $\Wit(x,y,w)$, $C$ accepts $(x,y,w)$, and if $\Pre(x)$ and
    $\neg \Wit(x,y,w)$, $C$ rejects the triple. 
  \end{enumerate}
\item[Witness Property:]
  We need to prove implication~(\ref{witness-property}).
\end{description}

In case of the maximum cardinality matching problem, the witness property states that an odd-set cover $\OSC$ as defined in Figure~\ref{Matching} with $\abs{M} = n_1 + \sum_{i \ge 2} \floor{n_i/2}$ proves that the matching $M$ has maximum cardinality. Checker correctness amounts to the statement that the program shown in Figure~\ref{Checker} is correct.

We proved the witness property using Isabelle/HOL~\cite{nipkow2002isabelle}; see Section~\ref{sec:Isabelle} for more information on Isabelle/HOL.
For the checker correctness, we used VCC~\cite{Cohen:TPHOLs2009-23} and later Simpl~\cite{schirmer:2006:thesis} and AutoCorres~\cite{greenaway-etal:2012:gap}. The latter approach has the advantage that the entire verification can be performed within Isabelle. Simpl is
a generic imperative programming language embedded into Isabelle/HOL, which was
designed as an intermediate language for program verification. We implemented checkers both in Simpl and C. Checkers written in Simpl were verified directly within Isabelle. For the checkers written in C, we first translated from C to Isabelle using the C-to-Isabelle
parser that was developed as part of the seL4 project~\cite{klein-etal:2010:sel4}, and then used the AutoCorres tool developed at NICTA that simplifies reasoning about C in Isabelle/HOL. Christine spent several months at NICTA to learn how to use the tool.
We verified the checkers for connectivity, maximum cardinality matching, and non-planarity. In particular, for the non-planarity checker it was essential that Lars Noschinski in parallel formalized basic graph theory in Isabelle~\cite{noschinski:2011:graph-library}.

A disclaimer is in order here. We did not verify the \CC\ program shown in~Figure~\ref{Checker}. Rather we verified a manual translation of this program into Simple or C, respectively. For this translation, we assumed a very basic representation of graphs. The nodes are numbered from 0 to $n-1$, the edges are numbered from $0$ to $m-1$ with the edges incident to any vertex numbered consecutively and arrays of the appropriate dimension are used for cross-referencing and for encoding adjacency lists.

The verification attempt for the maximum cardinality checker shown in Figure~\ref{Checker} discovered a flaw. Note that the program does not check whether the edges in $M$ actually belong to $G$. When we wrote the checker, we apparently took this for granted. The verification attempt revealed the flaw.

We also considered going further and briefly tried to verify the LEDA maximum cardinality matching algorithm~\cite[Section 7.7]{LEDAbook}. The program has 330 lines of code and the description of the algorithm, its implementation and its correctness proof spans over 20 pages. We found the task too daunting and, extrapolating from the effort required for the verification of the checkers, estimated the effort as several man-years.

{
\section{Level Four of Trustworthiness: Formal Verification of Complex Algorithms}

\providecommand{\insts}{}
\renewcommand{\insts}{\ensuremath{\Delta}}
\providecommand{\inst}{\ensuremath{\tvsal}}
\newcommand{\act}{\ensuremath{\pi}}
\newcommand{\asarrow}[1]{\vec{#1}}
\renewcommand{\vec}[1]{\overset{\rightarrow}{#1}}
\newcommand{\as}{\ensuremath{\vec{{\act}}}}

\newcommand{\etc}{\textit{etc.}}
\newcommand{\versus}{\textit{vs.}}
\renewcommand{\ie}{i.e.}
\newcommand{\Ie}{I.e.}
\newcommand{\eg}{e.g.}
\newcommand{\abziz}[1]{\textcolor{brown}{#1}}
\newcommand{\sublist}[2]{ \ensuremath{#1} \preceq\!\!\!\raisebox{.4mm}{\ensuremath{\cdot}}\; \ensuremath{#2}}
\newcommand{\subscriptsublist}[2]{\ensuremath{#1}\preceq\!\raisebox{.05mm}{\ensuremath{\cdot}}\ensuremath{#2}}
\newcommand{\PLS}{\Pi^\preceq\!\raisebox{1mm}{\ensuremath{\cdot}}}
\newcommand{\PLScharles}{\Pi^d}
\newcommand{\execname}{\mathsf{ex}}
\newcommand{\IndHyp}{\mathsf{IH}}
\newcommand{\exec}[2]{#2(#1)}

\newcommand{\ancestorssymbol}{\textsf{\upshape ancestors}}
\newcommand{\ancestors}{\ancestorssymbol}
\newcommand{\satpreas}[2]{\ensuremath{sat_precond_as(s, \as)}}
\newcommand{\proj}[2]{\ensuremath{#1{\downharpoonright}_{#2}}}
\newcommand{\dep}[3]{\ensuremath{#2 {\rightarrow} #3}}
\newcommand{\deptc}[3]{\ensuremath{#2 {\rightarrow^+} #3}}
\newcommand{\negdep}[3]{\ensuremath{#2 \not\rightarrow #3}}
\newcommand{\leavessymbol}{\textsf{\upshape leaves}}
\newcommand{\leaves}{\leavessymbol}

\newcommand{\childrensymbol}{\textsf{\upshape children}}
\newcommand{\children}[2]{\mathcal{\childrensymbol}_{#2}(#1)}
\newcommand{\succsymbol}{\textsf{\upshape succ}}
\newcommand{\succstates}[2]{\succsymbol(#1, #2)}
\newcommand{\concat}{\#}
\newcommand{\RG}{\cite{Rintanen:Gretton:2013}\ }
\newcommand{\cupdot}{\charfusion[\mathbin]{\cup}{\cdot}}
\newcommand{\bigcupdot}{\charfusion[\mathop]{\bigcup}{\cdot}}
\newcommand{\cuparrow}{\charfusion[\mathbin]{\cup}{{\raisebox{.5ex} {\smathcalebox{.4}{\ensuremath{\leftarrow}}}}}}
\newcommand{\bigcuparrow}{\charfusion[\mathop]{\bigcup}{\leftarrow}}
\newcommand{\finiteunion}{\cuparrow}
\newcommand{\finitemap}{\ensuremath{\sqsubseteq}}
\newcommand{\dgraph}{dependency graph}
\newcommand{\domain}[1]{{\sc #1}}
\newcommand{\solver}[1]{{\sc #1}}
\providecommand{\problem}[1]{\domain{#1}}
\renewcommand{\v}{\ensuremath{\mathit{v}}}
\providecommand{\vs}[1]{\domain{#1}}
\renewcommand{\vs}{\ensuremath{\mathit{vs}}}
\newcommand{\VS}{\ensuremath{\mathit{VS}}}
\newcommand{\Aut}{\ensuremath{\mathit{Aut}}}
\newcommand{\Inst}[2]{\ensuremath{\mathit{#2 \rightarrow_{#1} #1}}}
\newcommand{\Image}{\ensuremath{\mathit{Im}}}
\newcommand{\Img}[2]{\protect{#1 \llparenthesis #2 \rrparenthesis}}
\newcommand{\SND}{\ensuremath{\mathit{\pi_2}}}
\newcommand{\FST}{\ensuremath{\mathit{\pi_1}}}
\newcommand{\tvsal}{{\pitchfork}}
\newcommand{\nauty}{CGIP}

\newcommand{\pwinter}{\ensuremath{\mathit{\bigcap_{pw}}}}

\newcommand{\dom}{\ensuremath{\mathit{\mathcal{D}}}}
\newcommand{\codom}{\ensuremath{\mathcal{R}}}

\newcommand{\map}{\ensuremath{\mathit{map}}}
\newcommand{\BIJEC}{\ensuremath{\mathit{bij}}}
\newcommand{\INJ}{\ensuremath{\mathit{inj}}}
\newcommand{\funion}{\ensuremath{\overset{\leftarrow}{\cup}}}

\newcommand{\ifnew}{\mbox{\upshape \textsf{if}}}
\newcommand{\thennew}{\mbox{\upshape \textsf{then}}}
\newcommand{\elsenew}{\mbox{\upshape \textsf{else}}}
\newcommand{\choice}{\mbox{\upshape \textsf{ch}}}
\newcommand{\arbchoice}{\mbox{\upshape \textsf{arb}}}
\newcommand{\acycchoice}{\mbox{\upshape \textsf{ac}}}
\newcommand{\cycchoice}{\mbox{\upshape \textsf{cyc}}}
\newcommand{\filter}{\ensuremath{\mathit{FIL}}}
\newcommand{\probset}{\ensuremath{\boldsymbol \Pi}}
\newcommand{\probleq}{\ensuremath{\leq_\Pi}}
\newcommand{\CommVar}{\ensuremath{\bigcap_\v} }
\newcommand{\quotfun}{\ensuremath{ \mathcal{Q}}}

\newcommand{\apre}{\mbox{\upshape \textsf{pre}}}
\newcommand{\aeff}{\mbox{\upshape \textsf{eff}}}
\newcommand{\problist}{\ensuremath \probset}
\newcommand{\cat}{{\frown}}
\newcommand{\probproj}[2]{{#1}{\downharpoonright}^{#2}}
\newcommand{\preced}{\mathbin{\rotatebox[origin=c]{180}{\ensuremath{\rhd}}}}
\newcommand{\perm}{\ensuremath{\sigma}}
\newcommand{\invariant}[2]{\ensuremath{\mathit{inv({#1},{#2})}}}
\newcommand{\invstates}[1]{\ensuremath{\mathit{inv({#1})}}}
\newcommand{\probss}[1]{{\mathcal S}(#1)}
\newcommand{\parChildRel}[3]{\ensuremath{\negdep{#1}{#2}{#3}}}
\newcommand{\asessymbol}{\ensuremath{\mathbb{A}}}
\newcommand{\ases}[1]{{#1}^*}
\newcommand{\uniStates}{\ensuremath{\mathbb{U}}}
\newcommand{\recurrenceDiam}{\ensuremath{\mathit{rd}}}
\newcommand{\recurrenceAcycDiamfun}{\ensuremath{\mathit{{\mathfrak A}}}}
\newcommand{\recurrenceDiamfun}{\ensuremath{\mathit{\mathfrak R}}}
\newcommand{\traversalDiam}{\ensuremath{\mathit{td}}}
\newcommand{\traversalDiamfun}{\ensuremath{\mathit{\mathfrak T}}}
\newcommand{\isPrefix}[2]{\ensuremath{#1 \preceq #2}}
\providecommand{\path}{\ensuremath{\gamma}}
\newcommand{\aspath}{\ensuremath{\vec{\path}}}
\renewcommand{\path}{\ensuremath{\gamma}}
\newcommand{\n}{\textsf{\upshape n}}
\providecommand{\graph}{}
\renewcommand{\graph}{{\cal G}}
\newcommand{\undirgraph}{{\cal G}}
\renewcommand{\sset}{\ensuremath{\mbox{\upshape \textsf{ss}}}}
\renewcommand{\ss}{\ensuremath{\state s}}
\newcommand{\slist}{\ensuremath{\vec{\mbox{\upshape \textsf{ss}}}}}
\newcommand{\sll}{\ensuremath{\vec{\state}}}
\newcommand{\listset}{\mbox{\upshape \textsf{set}}}
\newcommand{\asset}{\ensuremath{\mathit{K}}}
\newcommand{\aslist}{\ensuremath{\mathit{\overset{\rightarrow}{\gamma}}}}
\newcommand{\head}{\mbox{\upshape \textsf{first}}}
\renewcommand{\max}{\textsf{\upshape max}}
\newcommand{\argmax}{\textsf{\upshape argmax}}
\renewcommand{\min}{\textsf{\upshape min}}
\newcommand{\bool}{\mbox{\upshape \textsf{bool}}}
\newcommand{\last}{\mbox{\upshape \textsf{last}}}
\newcommand{\front}{\mbox{\upshape \textsf{front}}}
\newcommand{\rot}{\mbox{\upshape \textsf{rot}}}
\newcommand{\stuff}{\mbox{\upshape \textsf{intlv}}}
\newcommand{\tail}{\mbox{\upshape \textsf{tail}}}
\newcommand{\ngrtoas}{\ensuremath{\mathit{\as_{\graph_\mathbb{N}}}}}
\newcommand{\vsfun}{\mbox{\upshape \textsf{vs}}}
\newcommand{\inits}{\mbox{\upshape \textsf{init}}}
\newcommand{\satprecondas}{\mbox{\upshape \textsf{sat-pre}}}
\newcommand{\remcondlessact}{\mbox{\upshape \textsf{rem-condless}}}
\providecommand{\state}{}
\renewcommand{\state}{x}
\newcommand{\statea}{x}
\newcommand{\stateb}{y}
\newcommand{\statec}{z}
\newcommand{\fals}{\mbox{\upshape \textsf{F}}}
\newcommand{\indices}{\ensuremath{V}}
\newcommand{\edges}{\ensuremath{E}}
\newcommand{\vertices}{\ensuremath{V}}
\newcommand{\listtype}{\mbox{\upshape \textsf{list}}}
\newcommand{\settype}{\mbox{\upshape \textsf{set}}}
\newcommand{\acttype}{\mbox{\upshape \textsf{action}}}
\newcommand{\graphtype}{\mbox{\upshape \textsf{graph}}}
\newcommand{\projfun}[2]{\ensuremath{\Delta_{#1}^{#2}}}
\newcommand{\snapfun}[2]{\ensuremath{\Sigma_{#1}^{#2}}}
\newcommand{\RDfun}[1]{\ensuremath{{\mathcal R}_{#1}}}
\newcommand{\elldbound}[1]{\ensuremath{{\mathcal LS}_{#1}}}
\newcommand{\distinct}{\textsf{\upshape distinct}}
\newcommand{\ddistinct}{\mbox{\upshape \textsf{ddistinct}}}
\newcommand{\simple}{\mbox{\upshape \textsf{simple}}}

\newcommand{\reachable}[3]{\ensuremath{{#1}\rightsquigarrow{#3}}}

\newcommand{\Omit}[1]{}

\newcommand{\charles}[1]{\textcolor{red}{#1}}

\newcommand{\negreachable}[3]{\ensuremath{{#2}\not\rightsquigarrow{#3}}}
\newcommand{\wdiam}[2]{{#1}^{#2}}
\newcommand{\dsnapshot}[2]{\Delta_{#1}}
\newcommand{\ellsnapshot}[2]{{\mathcal L}_{#1}}

\newcommand{\snapshotsymbol}{|\kern-.7ex\raise.08ex\hbox{\scalebox{0.7}{$\bullet$}}}
\newcommand{\snapshot}[2]{\ensuremath{\mathrel{#1\snapshotsymbol_{#2}}}}
\newcommand{\vstype}{\texttt{\upshape VS}}
\newcommand{\vtype}{{\scriptsize \ensuremath{\dom(\delta)}}}
\newcommand{\Balgo}{{\mbox{\textsc{Hyb}}}}
\newcommand{\ssgraph}[1]{\graph_\ss}
\newcommand{\agree}{\textsf{\upshape agree}}
\newcommand{\ck}{\ensuremath{\texttt{ck}}}
\newcommand{\lk}{\ensuremath{\texttt{lk}}}
\newcommand{\gr}{\ensuremath{\texttt{gr}}}
\newcommand{\gk}{\ensuremath{\texttt{gk}}}
\newcommand{\CK}{\ensuremath{\texttt{CK}}}
\newcommand{\LK}{\ensuremath{\texttt{LK}}}
\newcommand{\GR}{\ensuremath{\texttt{GR}}}
\newcommand{\GK}{\ensuremath{\texttt{GK}}}
\newcommand{\safe}{\ensuremath{\texttt{s}}}

\newcommand{\derivname}{\ensuremath{\partial}}
\newcommand{\deriv}[3]{\ensuremath{\derivname(#1,#2,#3)}}
\newcommand{\derivabbrev}[3]{\ensuremath{{\partial(#1,#2)}}}
\newcommand{\subsetoracle}{\ensuremath{ \Omega}}
\newcommand{\Aalgo}{{\mbox{\textsc{Pur}}}}
\newcommand{\Sname}{\textsf{\upshape S}}
\newcommand{\Sbrace}[1]{\Sname\langle#1\rangle}
\newcommand{\SalgoName}{\Sname_{\textsf{\upshape max}}}
\newcommand{\Salgo}[1]{\SalgoName\langle#1\rangle}

\newcommand{\WLPname}{{\mbox{\textsc{wlp}}}}
\newcommand{\WLPbrace}[1]{\WLPname\langle#1\rangle}
\newcommand{\WLPalgoName}{\WLPname_{\textsf{\upshape max}}}
\newcommand{\WLP}[1]{\WLPalgoName\langle#1\rangle}

\newcommand{\Nname}{\ensuremath{\textsf{\upshape N}}}
\newcommand{\Nbrace}[1]{\Nname\langle#1\rangle}
\newcommand{\NalgoName}{\Nname{_{\textsf{\upshape sum}}}}
\newcommand{\Nalgobrace}[1]{\NalgoName\langle#1\rangle}

\newcommand{\acycNname}{\widehat{\textsf{\upshape N}}}
\newcommand{\acycNbrace}[1]{\acycNname\langle#1\rangle}
\newcommand{\acycNalgoName}{\acycNname{_{\textsf{\upshape sum}}}}
\newcommand{\acycNalgobrace}[1]{\acycNalgoName\langle#1\rangle}

\newcommand{\Mname}{\ensuremath{\textsf{\upshape M}}}
\newcommand{\Mbrace}[1]{\Mname\langle#1\rangle}
\newcommand{\MalgoName}{\Mname{_{\textsf{\upshape sum}}}}
\newcommand{\Malgobrace}[1]{\MalgoName\langle#1\rangle}
\newcommand{\cardinality}[1]{{\ensuremath{|#1|}}}
\newcommand{\length}[1]{\cardinality{#1}}
\newcommand{\basecasefun}{\ensuremath{b}}
\newcommand{\Basecasefun}{\ensuremath{\mathcal B}}

\newcommand{\vertexgen}{\ensuremath{u}}
\newcommand{\vertexa}{{\ensuremath{\vertexgen_1}}}
\newcommand{\vertexb}{{\ensuremath{\vertexgen_2}}}
\newcommand{\vertexc}{{\ensuremath{\vertexgen_3}}}
\newcommand{\vertexd}{{\ensuremath{\vertexgen_4}}}
\newcommand{\vertexsetgen}{\ensuremath{\mathit{us}}}
\newcommand{\vertexseta}{\vertexsetgen_1}
\newcommand{\vertexsetb}{\vertexsetgen_2}
\newcommand{\labelsymbol}{\ensuremath{l}}
\newcommand{\labelfun}{\ensuremath{\mathcal{L}}}
\newcommand{\DAG}{\ensuremath{A}}
\newcommand{\NalgoNameN}{{\ensuremath{\NalgoName_{\mathbb{N}}}}}
\newcommand{\NnameN}{\ensuremath{\Nname_\mathbb{N}}}
\newcommand{\replaceprojsinglename}{\raisebox{-0.3mm} {\scalebox{0.7}{\textpmhg{H}}}}
\newcommand{\replaceprojsingle}[3] {{ #2} \underset {#1} {\raisebox{-0.3mm} {\scalebox{0.7}{\textpmhg{H}}}}  #3}
\newcommand{\HOLreplaceprojsingle}[1]{\underset {#1} {\raisebox{-0.3mm} {\scalebox{0.7}{\textpmhg{H}}}}}

\newcommand{\lotus}{{\scalebox{0.6}{\includegraphics{lotus.pdf}}}}
\newcommand{\invlotus}{\mathbin{\rotatebox[origin=c]{180}{$\lotus$}}}
\newcommand{\clique}{\ensuremath{K}}
\newcommand{\partition}{\ensuremath{\vs_{1..n}}}
\newcommand{\partitiontype}{\ensuremath{\vstype_{1..n}}}
\newcommand{\vtxpartition}{\ensuremath{P}}

\newcommand{\traversalDiamAlgo}{{\mbox{\textsc{TravDiam}}}}
\newcommand{\prefix}{\textsf{\upshape pfx}}
\newcommand{\powerset}{\mathbb{P}}
\newcommand{\postfix}{\textsf{\upshape sfx}}
\newcommand{\dfunproj}{\ensuremath{{\mathfrak D}}}
\newcommand{\dfunsnap}{\ensuremath{{\textgoth D}}}
\newcommand{\ellfunproj}{\ensuremath{\mathfrak L}}
\newcommand{\ellfunsnap}{\ensuremath{\textgoth L}}
\newcommand{\cycle}{\ensuremath{C}}
\newcommand{\petal}{\ensuremath{\eta}}
\renewcommand{\prod}{\ensuremath{{{{{\mathlarger{\mathlarger {{\mathlarger {\Pi}}}}}}}}}}
\newcommand{\sccset}{{\ensuremath{SCC}}}
\newcommand{\scc}{{\ensuremath{scc}}}
\newcommand{\negate}[1]{\overline{#1}}
\newcommand{\setofsets}{\ensuremath{S}}
\newcommand{\group}{\ensuremath{\cal \Gamma}}
\newcommand{\neededvars}{{\cal N}}
\newcommand{\sspace}{\mbox{\upshape \textsf{sspc}}}
\newcommand{\tip}{\ensuremath{t}}
\newcommand{\vara}{\ensuremath{\v_1}}
\newcommand{\varb}{\ensuremath{\v_2}}
\newcommand{\varc}{\ensuremath{\v_3}}
\newcommand{\vard}{\ensuremath{\v_4}}
\newcommand{\vare}{\ensuremath{\v_5}}
\newcommand{\varf}{\ensuremath{\v_6}}
\newcommand{\varg}{\ensuremath{\v_7}}
\newcommand{\varh}{\ensuremath{\v_8}}
\newcommand{\vari}{\ensuremath{\v_9}}
\newcommand{\acta}{\ensuremath{\act_1}}
\newcommand{\actb}{\ensuremath{\act_2}}
\newcommand{\actc}{\ensuremath{\act_3}}
\newcommand{\actd}{\ensuremath{\act_4}}
\newcommand{\acte}{\ensuremath{\act_5}}
\newcommand{\actf}{\ensuremath{\act_6}}
\newcommand{\actg}{\ensuremath{\act_7}}
\newcommand{\acth}{\ensuremath{\act_8}}
\newcommand{\acti}{\ensuremath{\act_9}}

\tikzset{dots/.style args={#1per #2}{line cap=round,dash pattern=on 0 off #2/#1}}
\providecommand{\moham}[1]{\fbox{{\bf \@Mohammad: }#1}}
\newcommand{\TDbound}{{\mbox{\textsc{Arb}}}}
\newcommand{\expbound}{{\mbox{\textsc{Exp}}}}
\newcommand{\sasdom}{\expbound}
\newcommand{\cardfun}{\ensuremath{\mathbb{C}}}
\newcommand{\AGNa}{AGN1}
\newcommand{\AGNb}{AGN2}
\newcommand{\reset}{{\ensuremath{reset}}}

\newcommand{\matching}{{\cal M}}
\newcommand{\BlossomAlg}{{\mbox{\textsc{Find\_Max\_Matching}}}}
\newcommand{\AugPathAlg}{{\mbox{\textsc{Aug\_Path\_Search}}}}
\newcommand{\BlossomOrAugPath}{{\mbox{\textsc{Compute\_Blossom}}}}

A decade later, we perform the formal verification of the blossom-shrinking algorithm for maximum cardinality. We give a short account of the verification which will be described in detail in our forthcoming publication~\cite{BlossomCorrectnessProof}.
On a high-level Edmond's blossom-shrinking algorithm~\cite{Edmonds:matching} works as follows. The algorithm repeatedly searches for an augmenting path with respect to the current matching. Initially, the current matching is empty. Whenever an augmenting path is found, augmentation of the path increases the size of the matching by one. If no augmenting path exists with respect to the current matching, the current matching has maximum cardinality.

The search for an augmenting path is via growing alternating trees rooted at free vertices, i.e. vertices not incident to an edge of the matching. The search is initialised by making each free vertex a root of an alternating tree; the matched nodes are in no tree initially. In an alternating tree, vertices at even depth are entered by a matching edge, vertices at odd depth are entered by a non-matching edge, and all leaves have even depth. In each step of the growth process, one considers a vertex, say $\vertexa$, of even depth that is incident to an edge $\{\vertexa,\vertexb\}$ not considered before. If $\vertexb$ is not in a tree yet, then one adds $\vertexb$ (at odd level) and its mate (at even level) under the current matching to the tree. If $\vertexb$ is already in a tree and has odd level then one does nothing as one simply has discovered another odd length path to $\vertexb$. If $\vertexb$ is already in a tree and has even level then one has either discovered an augmenting path (if $\vertexb$ is in a different tree than $\vertexa$) or a blossom (if $\vertexb$ and $\vertexa$ are in the same tree). In the latter case, consider the tree paths from $\vertexb$ and $\vertexa$ back to their common root and let $\vertexc$ be the lowest common ancestor of $\vertexb$ and $\vertexa$. The edge $\{\vertexa,\vertexb\}$ plus the tree paths from $\vertexa$ and $\vertexb$ to $\vertexc$ form an odd length cycle. One collapses all nodes on the cycle into a single node and repeats the search for an augmenting path in the quotient (= shrunken) graph. If an augmenting path is found in the quotient graph, it is lifted (refined) to an augmenting path in the original graph. If no augmenting path exists in the quotient graph, no augmenting path exists in the original graph.
In this section, we describe in detail the algorithm outlined above, and the process of formalising and verifying it in Isabelle/HOL.

\subsection{Isabelle/HOL}
\label{sec:Isabelle}
Isabelle/HOL~\cite{paulson1994isabelle} is a theorem prover for classical Higher-Order Logic. 
Roughly speaking, Higher-Order Logic can be seen as a combination of functional programming with logic.
Isabelle's syntax is a variation of Standard ML combined with (almost) standard mathematical notation.
Application of function $f$ to arguments $x_1~\ldots~x_n$ is written as $f~x_1~\ldots~x_n$ instead of the standard notation $f(x_1,~\ldots~,x_n)$.
We explain non-standard syntax in the paper where it occurs.

Isabelle is designed for trustworthiness: following the LCF approach~\cite{milner1972logic}, a small kernel implements the inference rules of the logic, and, using encapsulation features of ML, it guarantees that all theorems are actually proved by this small kernel.
Around the kernel, there is a large set of tools that implement proof tactics and high-level concepts like algebraic data types and recursive functions.
Bugs in these tools cannot lead to inconsistent theorems being proved since they all rely
on the kernel only, but only to error messages when the kernel refuses a proof.
Isabelle/HOL comes with a rich set of already
formalized theories, among which are natural numbers and integers as well
as sets and finite sets.

\subsection{Preliminaries}
An edge is a set of vertices with size 2.
A graph $\graph$ is a set of edges.
A set of edges $\matching$ is a matching iff $\forall e,e'\in\matching.\; e \cap e' = \emptyset$.
In Isabelle/HOL that is represented as follows:
\begin{isabelle}
\ \ matching\ M\ {\isasymlongleftrightarrow}\ ({\isasymforall}e{\isadigit{1}}\ {\isasymin}\ M.\ {\isasymforall}e{\isadigit{2}}\ {\isasymin}\ M.\ e{\isadigit{1}}\ {\isasymnoteq}\ e{\isadigit{2}}\ {\isasymlongrightarrow}\ e{\isadigit{1}}\ {\isasyminter}\ e{\isadigit{2}}\ =\ \{\})
\end{isabelle}
In may cases, a matching is a subset of a graph, in which case we call it a matching w.r.t. the graph.
For a graph $\graph$, $\matching$ is a maximum matching w.r.t.\ $\graph$ iff for any matching $\matching'$ w.r.t. $\graph$, we have that $\cardinality{\matching'} \leq  \cardinality{\matching}$.

\subsection{Formalising Berge's Lemma}
\label{sec:berge}

A list of vertices $\vertexa\vertexb\dots\vertexgen_n$ is a path w.r.t.\ a graph $\graph$ iff every $\{\vertexgen_i, \vertexgen_{i+1}\}\in\graph$.
A path $\vertexa\vertexb\dots\vertexgen_n$ is a simple path iff for every $1\leq i\neq j\leq n$, $\vertexgen_i \neq \vertexgen_j$.
A list of vertices $\vertexa\vertexb\dots\vertexgen_n$ is an alternating path w.r.t.\ a set of edges $\edges$ iff for some $\edges'$ (1) $\edges' = \edges$ or $\edges' = \{e\mid e \not\in\edges\}$, (2) $\{\vertexgen_i,\vertexgen_{i+1}\}\in\edges'$ holds for all even numbers $i$, where $1 \leq i < n$, and (3) $\{\vertexgen_i,\vertexgen_{i+1}\}\not\in\edges'$ holds for all odd numbers $i$, where $1 \leq i \leq n$.
We call a list of vertices $\vertexa\vertexb\dots\vertexgen_n$ an augmenting path w.r.t. a matching $\matching$ iff $\vertexa\vertexb\dots\vertexgen_n$ is an alternating path w.r.t. $\matching$ and $\vertexa,\vertexgen_n\not\in\bigcup\matching$. 
It is often the case that an augmenting path $\path$ w.r.t. to a matching $\matching$ is also a simple path w.r.t. a graph $\graph$, in which case we call the path an augmenting path w.r.t. to the pair $\langle\graph,\matching\rangle$.
Also, for two sets $s$ and $t$, $s \oplus t$ denotes the symmetric difference of the two sets.
We overload $\oplus$ to arguments which are lists in the obvious fashion.


\begin{mythm}[Berge's Lemma]
\label{thm:berge}
For a graph $\graph$, a matching $\matching$ is maximum w.r.t.\ $\graph$ iff there is not an augmenting path $\path$ w.r.t.\ $\langle \graph, \matching\rangle$.
\end{mythm}
Our proof of Berge's lemma is shorter than the standard proof. The standard proof consists of three steps.
First, for any two matchings $\matching$ and $\matching'$, every connected component of the graph $\matching \oplus \matching'$ is either a singleton vertex, a path, or a cycle. Second, for a set of edges $C \subseteq \matching \oplus \matching'$ s.t. $\cardinality{C \cap \matching} < \cardinality{C \cap \matching'}$, the edges from $C$ form a path.
Third, such a set $C$ of edges exists, if $\cardinality{\matching} < \cardinality{\matching'}$.
We observe that it is easier to directly show that such a $C$ exists and that all its edges can be arranged in a path, without having to prove the first step about all connected components.
We found this different proof during the process of formalising the theorem, and finding this shorter proof was primarily motivated by making the formalisation shorter and more feasible.
The discovery of simpler proofs or more general theorem statements is one potential positive outcome of verifying algorithms, and mathematics in general, in interactive theorem provers~\cite{DBLP:journals/jar/AbdulazizPaulson18,DBLP:journals/jar/AbdulazizNG18,dahmen2019formalizing}.



\SetKwIF{If}{ElseIf}{Else}{if}{}{else if}{else}{endif}
\begin{algorithm}\DontPrintSemicolon
  $\path := \AugPathAlg(\graph,\matching)$\;
  \KwSty{if} $\gamma$ {\upshape is some augmenting path}
    \\ \ \ \KwSty{return} {$\BlossomAlg(\graph,\matching\oplus\path)$}

  \KwSty{else}
    \\ \ \ \KwSty{return} {$\matching$}
  \caption{$\BlossomAlg(\graph, \matching)$}\label{alg:Blossom}
\end{algorithm}

Now consider Algorithm~\ref{alg:Blossom}.
Berge's lemma implies the validity of that algorithm as a method to compute maximum matchings in graphs.
The validity of Algorithm~\ref{alg:Blossom} is stated in the following corollary.
\begin{mycor}
\label{cor:BlossomWorks}
Assume that $\AugPathAlg(\graph,\matching)$ is an augmenting path w.r.t.\ $\langle\graph,\matching\rangle$, for any graph $\graph$ and matching $\matching$, iff $\graph$ has an augmenting path w.r.t.\ $\langle\graph,\matching\rangle$.
Then, for any graph $\graph$, $\BlossomAlg(\graph, \emptyset)$ is a maximum matching w.r.t.\ $\graph$.
\end{mycor}
As shown in Corollary~\ref{cor:BlossomWorks}, Algorithm~\ref{alg:Blossom} depends on the function $\AugPathAlg$ which is a sound and a complete procedure to compute augmenting paths in graphs.

In Isabelle/HOL, the first step is to formalise the path concepts from above.
Paths and alternating paths are defined recursively in a straightforward fashion.
An augmenting path is defined as follows:
\begin{isabelle}
\ \ augmenting\_path\ M\ p\ {\isasymequiv}\ (length\ p\ {\isasymge}\ {\isadigit{2}})\ {\isasymand}\ alt\_path\ M\ p\isanewline
\ \ \ \ \ \ \ \ \ \ \ \ \ \ \ \ \ \ \ \ \ \ \ \ \ \ {\isasymand}\ hd\ p\ {\isasymnotin}\ Vs\ M\ {\isasymand}\ last\ p\ {\isasymnotin}\ Vs\ M
\end{isabelle}
The formalised statement of Berge's lemma is as follows:
\begin{isabelle}
\isacommand{theorem}\isamarkupfalse%
\ Berge:\isanewline
\ \ \isakeyword{assumes}\isanewline
\ \ \ \ finite\ M\ \isakeyword{and}\ matching\ M\ \isakeyword{and}\ M\ {\isasymsubseteq}\ E\isanewline
\ \ \ \ \isakeyword{and}\isanewline
\ \ \ \ ({\isasymforall}e{\isasymin}E.\ {\isasymexists}u\ v.\ e\ =\ \{u,v\}\ {\isasymand}\ u\ {\isasymnoteq}\ v)\ \isakeyword{and}\ finite\ (Vs\ E)\isanewline
\ \ \isakeyword{shows}\ ({\isasymexists}p.\ augmenting\_path\ M\ p\ {\isasymand}\ path\ E\ p\ {\isasymand}\ distinct\ p)\ {\isasymlongleftrightarrow}\isanewline
\ \ \ \ \ \ \ \ \ \ \ \ ({\isasymexists}M{\isacharprime}\ {\isasymsubseteq}\ E.\ matching\ M{\isacharprime}\ {\isasymand}\ card\ M\ {\isacharless}\ card\ M{\isacharprime})
\end{isabelle}
Note that in the formalisation when the paths need to be simple, such as in Berge's lemma above, we have the additional assumption that all vertices are pairwise distinct, denoted by the Isabelle/HOL predicate \isa{distinct}.
Just to clarify Isabelle's syntax: the lemma above has two sets of assumptions, one on the matching and the other on the graph.
The matching has to be a finite set, which is a matching w.r.t. the given graph.
The graph has to have edges which only have two vertices, and its set of vertices has to be finite.

In Isabelle/HOL Algorithm~\ref{alg:Blossom} is formalised within the following \emph{locale}.
\begin{isabelle}
\isacommand{locale}\isamarkupfalse%
\ find\_max\_match\ =\isanewline
\ \ \isakeyword{fixes}\ aug\_path\_search::{\isacharprime}a\ set\ set\ {\isasymRightarrow}\ {\isacharprime}a\ set\ set\ {\isasymRightarrow}\ ({\isacharprime}a\ list)\ option\ \isakeyword{and}\isanewline
\ \ \ \ E\isanewline
\ \ \isakeyword{assumes}\isanewline
\ \ \ \ aug\_path\_search\_complete:\ \isanewline
\ \ \ \ matching\ M\ {\isasymand}\ M\ {\isasymsubseteq}\ E\ {\isasymand}\ finite\ M\ {\isasymand}\isanewline
\ \ \ \ \ \ ({\isasymexists}p.\ path\ E\ p\ {\isasymand}\ distinct\ p\ {\isasymand}\ augmenting\_path\ M\ p)\isanewline
\ \ \ \ \ \ \ \ \ {\isasymLongrightarrow}\ ({\isasymexists}p.\ aug\_path\_search\ E\ M\ =\ Some\ p)\isanewline
\ \ \ \ \isakeyword{and}\isanewline
\ \ \ \ aug\_path\_search\_sound:\isanewline
\ \ \ \ matching\ M\ {\isasymand}\ M\ {\isasymsubseteq}\ E\ {\isasymand}\ finite\ M\ {\isasymand}\ aug\_path\_search\ E\ M\ =\ Some\ p\ {\isasymLongrightarrow}\isanewline
\ \ \ \ \ \ \ \ \ \ \ \ \ path\ E\ p\ {\isasymand}\ distinct\ p\ {\isasymand}\ augmenting\_path\ M\ p\isanewline
\ \ \ \ \isakeyword{and}\isanewline
\ \ \ \ graph:\ {\isasymforall}e{\isasymin}E.\ {\isasymexists}u\ v.\ e\ =\ \{u,\ v\}\ {\isasymand}\ u\ {\isasymnoteq}\ v\ finite\ (Vs\ E)
\end{isabelle}
A locale is a named context: definitions and theorems proved within locale \isa{find\_max\_match} can refer to the parameters and assumptions declared there.
In this case, we need the locale to identify the parameter \isa{aug\_path\_search} of the locale, corresponding to the function $\AugPathAlg$, which is used in Algorithm~\ref{alg:Blossom}.
The function \isa{aug\_path\_search} should take as input a graph and a matching.
It should return an \isa{('a list) option} typed value.
Generally speaking, the value of an \isa{'a option} valued term could be in one of two forms: either \isa{Some x}, or \isa{None}, where \isa{x} is of type \isa{'a}.
In the case of \isa{aug\_path\_search}, it should return either \isa{Some p}, where \isa{p} is a path in case an augmenting path is found, or \isa{None}, otherwise.
There is also the function \isa{the}, which given a term of type \isa{'a option}, returns \isa{x}, if the given term is \isa{Some x}, and which is undefined otherwise.
Within that locale, the definition of Algorithm~\ref{alg:Blossom} and its verification theorem are as follows.
Note that the verification theorem has four conclusions: the algorithm returns a subset of the graph, that subset is a matching, that matching is finite and the cardinality of any other matching is bounded by the size of the returned matching.
\begin{isabelle}
\ \ find\_max\_matching\ M\ =\ \isanewline
\ \ \ \ \ (if\ ({\isasymexists}p.\ aug\_path\_search\ E\ M\ =\ Some\ p)\ then\isanewline
\ \ \ \ \ \ \ \ (find\_max\_matching\ (M\ {\isasymoplus}\ (set\ (edges\_of\_path\ (the\ (aug\_path\_search\ E\ M))))))\isanewline
\ \ \ \ \ \ else\ M)\isanewline

\isacommand{lemma}\isamarkupfalse%
\ find\_max\_matching\_works:\isanewline
\ \ \isakeyword{shows}\ (find\_max\_matching\ \{\})\ {\isasymsubseteq}\ E\isanewline
\ \ \ \ matching\ (find\_max\_matching\ \{\})\isanewline
\ \ \ \ finite\ (find\_max\_matching\ \{\})\isanewline
\ \ \ \ {\isasymforall}M.\ matching\ M\ {\isasymand}\ M\ {\isasymsubseteq}\ E\ {\isasymand}\ finite\ M\ {\isasymlongrightarrow}\ card\ M\ {\isasymle}\ card\ (find\_max\_matching\ \{\})\isanewline
\end{isabelle}

Functions defined within a locale are parameterised on the constants which are declared in the locale's definition.
When a function is used outside a locale, these parameters must be specified. So, if \isa{find\_max\_matching} is used outside the locale above, it should take a function which computes augmenting paths as a parameter.
Similarly, theorems proven within a locale implicitly have the assumptions of the locale. So if we use the lemma \isa{find\_max\_matching\_works}, we would have to prove that the functional argument to \isa{find\_max\_matching} satisfies the assumptions of the locale, i.e. that argument is a sound and complete procedure for computing augmenting paths.
The way theorems from locales are used will be clearer in the next section when we refer to the function \isa{find\_max\_matching} and use the lemma \isa{find\_max\_matching\_works} outside of the locale \isa{find\_max\_match}.
The use of locales for performing gradual refinement of algorithms allows to focus on the specific aspects of the algorithm relevant to a refinement stage, with the rest of the algorithm abstracted away.

\subsection{Verifying that Blossom Contraction Works}\label{subsec:Contraction}
In Corollary~\ref{cor:BlossomWorks}, which specifies the soundness of $\BlossomAlg$, we have not explicitly specified the function $\AugPathAlg$.
Indeed, we have only specified what its output has to conform to.
We now refine that specification and describe $\AugPathAlg$ algorithmically.

Firstly, for a function $f$ and a set $s$, let $\Img{f}{s}$ denote the image of $f$ on $s$.
Also, for a set of edges $\edges$, and a function $f$, the quotient $\edges/f$ is the set $\{\Img{f}{e}\mid e\in\edges\}$.
We now introduce the concepts of a \emph{blossom}.
A list of vertices $\vertexa\vertexb\dots\vertexgen_n$ is called a cycle if $3 < n$ and $\vertexgen_n = \vertexa$, and we call it an odd cycle if $n$ is even.
A pair $\langle \vertexa\vertexb\dots\vertexgen_{i-1}, \vertexgen_i\vertexgen_{i+1}\dots\vertexgen_n\rangle$ is a blossom w.r.t.\ a matching $\matching$ iff (1)  $\vertexgen_i\vertexgen_{i+1}\dots\vertexgen_n$ is an odd cycle, (2) $\vertexa\vertexb\dots\vertexgen_n$ is an alternating path w.r.t.\ $\matching$, and (3) $\vertexa\not\in\bigcup\matching$.
We also refer to $\vertexa\vertexb\dots\vertexgen_i$ as the stem of the blossom.
In many situations we have a pair $\langle \vertexa\vertexb\dots\vertexgen_{i-1}, \vertexgen_i\vertexgen_{i+1}\dots\vertexgen_n\rangle$ which is a blossom w.r.t.\ a matching $\matching$ where $\vertexa\vertexb\dots\vertexgen_{i-1} \vertexgen_i\vertexgen_{i+1}\dots\vertexgen_{n-1}$ is also a simple path w.r.t. a graph $\graph$ and $\{\vertexgen_{n-1},\vertexgen_n\}\in\graph$.
In this case we call it a blossom w.r.t.\ $\langle\graph,\matching\rangle$.

Based on the above definitions, we prove that contracting (i.e. shrinking) the odd cycle of a blossom preserves the existence of an augmenting path, which is the second main result needed to prove the validity of the blossom-shrinking algorithm, after Berge's lemma.

\begin{mythm}
\label{thm:quotient}
Consider a graph $\graph$ and a vertex $u\not\in\bigcup\graph$.
Let for a set $s$, the function $P_s$ be defined as $P_s(x)= \ifnew\; x \in s\; \thennew\; u\; \elsenew\; x$.
Then, for a blossom $\langle \path, \cycle\rangle$ w.r.t. $\langle\graph,\matching\rangle$, if $s$ is the set of vertices in $\cycle$, then we have an augmenting path w.r.t.\ $\langle\graph,\matching\rangle$ iff there is an augmenting path w.r.t.\ $\langle\graph/P_s,\matching/P_s\rangle$.
\end{mythm}

\newcommand{\refine}{\textsf{\upshape refine}}
Theorem~\ref{thm:quotient} is used in most expositions of the blossom-shrinking algorithm.
In our proof for the forward direction (if an augmenting path exists w.r.t. $\langle\graph,\matching\rangle$, then there is an augmenting path w.r.t. $\langle\graph/P_s,\matching/P_s\rangle$, i.e. w.r.t. the quotients), we follow a standard textbook approach~\cite{Korte-Vygen}. In our proof for the backward direction (an augmenting path w.r.t. the quotients can be lifted to an augmenting path w.r.t. the original graph) we define an (almost) executable function $\refine$ that does the lifting.\footnote{The function $\refine$, as defined later, is executable except for a choice operation.}
We took the choice of explicitly defining that function with using it in the final algorithm in mind.
This is similar to the approach used in the informal proof of soundness of the variant of the blossom-shrinking algorithm used in LEDA~\cite{LEDAbook}.





\newcommand{\stem}{s}

Now, using Theorem~\ref{thm:quotient}, one can show that Algorithm~\ref{alg:AugPath} is a sound and complete procedure for computing augmenting paths.

\SetKwIF{If}{ElseIf}{Else}{if}{}{else if}{else}{endif}
\begin{algorithm}\DontPrintSemicolon
  \KwSty{if} $\BlossomOrAugPath(\graph, \matching)$ {\upshape is a blossom $\langle \path, \cycle \rangle$ w.r.t. $\langle\graph,\matching\rangle$}
    \\ \ \ \KwSty{return} {$\refine(\AugPathAlg(\graph/P_\cycle, \matching/P_\cycle))$}

  \KwSty{else} \KwSty{if} $\BlossomOrAugPath(\graph, \matching)$ {\upshape is an augmenting path w.r.t. $\langle\graph,\matching\rangle$}
    \\ \ \ \KwSty{return} {$\BlossomOrAugPath(\graph, \matching)$}

  \KwSty{else}
    \\ \ \ \KwSty{return} {no augmenting path found}
  \caption{$\AugPathAlg(\graph, \matching)$}\label{alg:AugPath}
\end{algorithm}

The soundness and completeness of this algorithm assumes that $\BlossomOrAugPath$ can successfully compute a blossom or an augmenting path in a graph iff either one exists.
This is formally stated as follows.
\begin{mycor}
\label{thm:AugPathAlgWorks}
Assume that, for a graph $\graph$ and a matching $\matching$ w.r.t.\ $\graph$, there is a blossom or an augmenting path w.r.t.\ $\langle\graph,\matching\rangle$ iff $\BlossomOrAugPath(\graph, \matching)$ is a blossom or an augmenting path w.r.t.\ $\langle\graph,\matching\rangle$.
Then for any graph $\graph$ and matching $\matching$, $\AugPathAlg(\graph, \matching)$ is an augmenting path w.r.t.\ $\langle\graph,\matching\rangle$ iff there is an augmenting path w.r.t.\ $\langle\graph,\matching\rangle$.
\end{mycor}

To formalise that in Isabelle/HOL, an odd cycle and a blossom are defined as follows:
\begin{isabelle}
\ \ odd\_cycle\ p\ {\isasymequiv}\ (length\ p\ {\isasymge}\ {\isadigit{3}})\ {\isasymand}\ odd\ (length\ (edges\_of\_path\ p))\ {\isasymand}\ hd\ p\ =\ last\ p\isanewline

\ \ blossom\ M\ stem\ C\ {\isasymequiv}\ alt\_path\ M\ (stem\ {\isacharat}\ C)\ {\isasymand}\ \isanewline
\ \ \ \ distinct\ (stem\ {\isacharat}\ (butlast\ C))\ {\isasymand}\ odd\_cycle\ C\ {\isasymand}\ hd\ (stem\ {\isacharat}\ C)\ {\isasymnotin}\ Vs\ M\ {\isasymand}\isanewline
\ \ \ \ even\ (length\ (edges\_of\_path\ (stem\ {\isacharat}\ {\isacharbrackleft}hd\ C{\isacharbrackright})))\isanewline
\end{isabelle}
In the above definition \isa{@} stands for list concatenation and \isa{edges\_of\_path} is a function which, given a path, returns the list of edges constituting the path.

To define the function $\refine$ that refines a quotient augmenting path to a concrete one or to formalise the theorems showing that contracting blossoms preserves augmenting paths we first declare the following locale:

\begin{isabelle}
\isacommand{locale}\isamarkupfalse%
\ quot\ =\isanewline
\ \ \isakeyword{fixes}\ P\ s\ u\isanewline
\ \ \isakeyword{assumes}\ {\isasymforall}v{\isasymin}s.\ P\ v\ =\ v\ \isakeyword{and} u{\isasymnotin}s\ \isakeyword{and}\ ({\isasymforall}v.\ v{\isasymnotin}s\ {\isasymlongrightarrow}\ P\ v\ =\ u)
\end{isabelle}
That locale fixes a function \isa{P}, a set of vertices \isa{s} and a vertex \isa{u}.
The function \isa{P} maps all vertices from \isa{s} to the given vertex \isa{u}.

Now, we formalise the function $\refine$ which lifts an augmenting path in a quotient graph to an augmenting path in the concrete graph. The function $\refine$ takes an augmenting path $p$ in the quotient graph and returns it unchanged if it does not contain the vertex $u$ and deletes $u$ and splits $p$ into two paths $p_1$ and $p_2$ otherwise. In the latter case, $p_1$ and $p_2$ are passed to \isa{replace\_cycle}. This function first defines two auxiliary paths \isa{stem2p2} and \isa{p12stem} using the function \isa{stem{\isadigit{2}}vert\_path}. Let us have a closer look at the path \isa{stem2p2}. \isa{stem{\isadigit{2}}vert\_path} with last argument \isa{hd p2} uses \isa{choose\_con\_vert} to find a neighbor of \isa{hd p2} on the cycle $C$. It splits the cycle at this neighbor and then returns the path leading to the base of the blossom starting with a matching edge. Finally, \isa{replace\_cycle} glues together $p_1$, $p_2$ and either \isa{stem2p2} and \isa{p12stem} to obtain an augmenting path in the concrete graph. 
\begin{isabelle}
\ \ choose\_con\_vert\ vs\ E\ v\ {\isasymequiv}\ (SOME\ v{\isacharprime}.\ v{\isacharprime}\ {\isasymin}\ vs\ {\isasymand}\ \{v,\ v{\isacharprime}\}\ {\isasymin}\ E)\isanewline

\ \ stem{\isadigit{2}}vert\_path\ C\ E\ M\ v\ {\isasymequiv}\isanewline
\ \ let\ find\_pfx{\isacharprime}\ =\ ({\isasymlambda}C.\ find\_pfx\ ((=)\ (choose\_con\_vert\ (set\ C)\ E\ v))\ C)\ in\isanewline
\ \ \ \ if\ (last\ (edges\_of\_path\ (find\_pfx{\isacharprime}\ C))\ {\isasymin}\ M)\ then\isanewline
\ \ \ \ \ \ (find\_pfx{\isacharprime}\ C)\isanewline
\ \ \ \ else\isanewline
\ \ \ \ \ \ (find\_pfx{\isacharprime}\ (rev\ C))\isanewline

\ \ replace\_cycle\ C\ E\ M\ p{\isadigit{1}}\ p{\isadigit{2}}\ {\isasymequiv}\isanewline
\ \ \ let\ stem{\isadigit{2}}p{\isadigit{2}}\ =\ stem{\isadigit{2}}vert\_path\ C\ E\ M\ (hd\ p{\isadigit{2}}){\isacharsemicolon}\isanewline
\ \ \ \ \ \ \ p{\isadigit{1}}{\isadigit{2}}stem\ =\ stem{\isadigit{2}}vert\_path\ C\ E\ M\ (last\ p{\isadigit{1}})\ in\isanewline
\ \ \ if\ p{\isadigit{1}}\ =\ {\isacharbrackleft}{\isacharbrackright}\ then\isanewline
\ \ \ \ \ stem{\isadigit{2}}p{\isadigit{2}}\ {\isacharat}\ p{\isadigit{2}}\isanewline
\ \ \ else\isanewline
\ \ \ \ \ (if\ p{\isadigit{2}}\ =\ {\isacharbrackleft}{\isacharbrackright}\ then\ \isanewline
\ \ \ \ \ \ \ \ p{\isadigit{1}}{\isadigit{2}}stem\ {\isacharat}\ (rev\ p{\isadigit{1}})\isanewline
\ \ \ \ \ \ else\isanewline
\ \ \ \ \ \ \ (if\ \{u,\ hd\ p{\isadigit{2}}\}\ {\isasymnotin}\ quotG\ M\ then\isanewline
\ \ \ \ \ \ \ \ \ p{\isadigit{1}}\ {\isacharat}\ stem{\isadigit{2}}p{\isadigit{2}}\ {\isacharat}\ p{\isadigit{2}}\isanewline
\ \ \ \ \ \ \ else\isanewline
\ \ \ \ \ \ \ \ \ (rev\ p{\isadigit{2}})\ {\isacharat}\ p{\isadigit{1}}{\isadigit{2}}stem\ {\isacharat}\ (rev\ p{\isadigit{1}})))\isanewline

\ \ refine\ C\ E\ M\ p\ {\isasymequiv}\isanewline
\ \ \ if\ (u\ {\isasymin}\ set\ p)\ then\isanewline
\ \ \ \ \ (replace\_cycle\ C\ E\ M\ (fst\ (pref\_suf\ {\isacharbrackleft}{\isacharbrackright}\ u\ p))\ (snd\ (pref\_suf\ {\isacharbrackleft}{\isacharbrackright}\ u\ p)))\isanewline
\ \ \ else\ p

\end{isabelle}

In Isabelle/HOL the two directions of the equivalence in Theorem~\ref{thm:quotient} are formalised as follows:

\begin{isabelle}

\isacommand{theorem}\isamarkupfalse%
\ quot\_apath\_to\_apath:\isanewline
\ \ \isakeyword{assumes}\isanewline
\ \ \ \  odd\_cycle\ C\ \isakeyword{and} alt\_path\ M\ C\ \isakeyword{and} distinct\ (tl\ C)\ \isakeyword{and} path\ E\ C\isanewline
\ \ \ \ \isakeyword{and}\isanewline
\ \ \ \ augmenting\_path\ (quotG\ M)\ p{\isacharprime}\ \isakeyword{and} distinct\ p{\isacharprime}\ \isakeyword{and} path\ (quotG\ E)\ p{\isacharprime}\isanewline
\ \ \ \ \isakeyword{and}\isanewline
\ \ \ \ matching\ M\ \isakeyword{and} M\ {\isasymsubseteq}\ E\isanewline
\ \ \ \ \isakeyword{and}\isanewline
\ \ \ \ s\ =\ (Vs\ E)\ -\ set\ C\isanewline
\ \ \ \ \isakeyword{and}\isanewline
\ \ \ \ {\isasymforall}e{\isasymin}E.\ {\isasymexists}u\ v.\ e\ =\ \{u,\ v\}\ {\isasymand}\ u\ {\isasymnoteq}\ v\isanewline
\ \ \isakeyword{shows}\ augmenting\_path\ M\ (refine\ C\ E\ M\ p{\isacharprime})\ {\isasymand}\ path\ E\ (refine\ C\ E\ M\ p{\isacharprime})\ {\isasymand}\isanewline
\ \ \ \ \ \ \ \ \ \ distinct\ (refine\ C\ E\ M\ p{\isacharprime})\isanewline

\isacommand{theorem}\isamarkupfalse%
\ aug\_path\_works\_in\_contraction:\isanewline
\ \ \isakeyword{assumes}\isanewline
\ \ \ \ path\ E\ (stem\ {\isacharat}\ C)\ \isakeyword{and} blossom\ M\ stem\ C\isanewline
\ \ \ \ \isakeyword{and}\isanewline
\ \ \ \ augmenting\_path\ M\ p\ \isakeyword{and} path\ E\ p\ \isakeyword{and} distinct\ p\isanewline
\ \ \ \ \isakeyword{and}\ \isanewline
\ \ \ \ matching\ M\ \isakeyword{and} M\ {\isasymsubseteq}\ E\ \isakeyword{and} finite\ M\isanewline
\ \ \ \ \isakeyword{and}\isanewline
\ \ \ \ s\ =\ (Vs\ E)\ -\ set\ C\ \isakeyword{and} u\ {\isasymnotin}\ \ Vs\ E\isanewline
\ \ \ \ \isakeyword{and}\isanewline
\ \ \ \ {\isasymforall}e{\isasymin}E.\ {\isasymexists}u\ v.\ e\ =\ \{u,\ v\}\ {\isasymand}\ u\ {\isasymnoteq}\ v\ \isakeyword{and} finite\ (Vs\ E)\isanewline
\ \ \isakeyword{shows}\ {\isasymexists}p{\isacharprime}.\ augmenting\_path\ (quotG\ M)\ p{\isacharprime}\ {\isasymand}\ path\ (quotG\ E)\ p{\isacharprime}\ {\isasymand}\ distinct\ p{\isacharprime}
\end{isabelle}

A main challenge with formalising Theorem~\ref{thm:quotient} in Isabelle/HOL is the lack of automation for handling symmetries in its proof.

To formalise Algorithm~\ref{alg:AugPath} we use a locale to assume the existence of the function which computes augmenting paths or blossoms, iff either one exist.
That function is called \isa{blos\_search} in the locale declaration.
Its return type and the assumptions on it are as follows:
\begin{isabelle}
\isacommand{datatype}\isamarkupfalse%
\ {\isacharprime}a\ blossom\_res\ =\isanewline
\ \ \ \  Path\ (aug\_path:\ "{\isacharprime}a\ list")\ {\isacharbar}\ Blossom\ (stem\_vs:\ "{\isacharprime}a\ list")\ (cycle\_vs:\ "{\isacharprime}a\ list")\isanewline

bloss\_algo\_complete:\ \isanewline
\ \ ((({\isasymexists}p.\ path\ E\ p\ {\isasymand}\ distinct\ p\ {\isasymand}\ augmenting\_path\ M\ p)\isanewline
\ \ \ \ \ \ \ \ \ {\isasymor}\ (matching\ M\ {\isasymand}\ ({\isasymexists}stem\ C.\ path\ E\ (stem\ {\isacharat}\ C)\ {\isasymand}\ blossom\ M\ stem\ C))))\isanewline
\ \ \ \ \ {\isasymLongrightarrow}\ ({\isasymexists}blos\_comp.\ blos\_search\ E\ M\ =\ Some\ blos\_comp)\isanewline

bloss\_algo\_sound:\isanewline
\ \ ({\isasymforall}e{\isasymin}E.\ {\isasymexists}u\ v.\ e\ =\ \{u,\ v\}\ {\isasymand}\ u\ {\isasymnoteq}\ v)\ {\isasymand}\ blos\_search\ E\ M\ =\ Some\ (Path\ p)\isanewline
\ \ \ \ \ \  {\isasymLongrightarrow}\ (path\ E\ p\ {\isasymand}\ distinct\ p\ {\isasymand}\ augmenting\_path\ M\ p)\isanewline
\ \ blos\_search\ E\ M\ =\ Some\ (Blossom\ stem\ C)\isanewline
\ \ \ \ \ \ {\isasymLongrightarrow}\ (path\ E\ (stem\ {\isacharat}\ C)\ {\isasymand}\ (matching\ M\ {\isasymlongrightarrow}\ blossom\ M\ stem\ C))
\end{isabelle}
The locale also fixes a function \isa{create\_vert} which creates new vertex names to which vertices from the odd cycle are mapped during contraction.
Within that locale, we define Algorithm~\ref{alg:AugPath} and prove its soundness and completeness theorems, which are as follows:
\begin{isabelle}

\ quotG\ E\ {\isasymequiv}\ (quot\_graph\ P\ E)\ -\ \{\{u\}\}\ \isanewline

\ find\_aug\_path\ E\ M\ =\ \isanewline
\ \ \ \ (case\ blos\_search\ E\ M\ of\ Some\ blossom\_res\ {\isasymRightarrow}\isanewline
\ \ \ \ \ \ \ case\ blossom\_res\ of\ Path\ p\ {\isasymRightarrow}\ Some\ p\isanewline
\ \ \ \ \ \ \ {\isacharbar}\ Blossom\ stem\ cyc\ {\isasymRightarrow}\isanewline
\ \ \ \ \ \ \ \ \ \ \ let\ u\ =\ create\_vert\ (Vs\ E){\isacharsemicolon}\isanewline
\ \ \ \ \ \ \ \ \ \ \ \ \ \ \ s\ =\ Vs\ E\ -\ (set\ cyc){\isacharsemicolon}\isanewline
\ \ \ \ \ \ \ \ \ \ \ \ \ \ \ quotG\ =\ quot.quotG\ (quot\_fun\ s\ u)\ u{\isacharsemicolon}\isanewline
\ \ \ \ \ \ \ \ \ \ \ \ \ \ \ refine\ =\ quot.refine\ (quot\_fun\ s\ u)\ u\ cyc\ E\ M\isanewline
\ \ \ \ \ \ \ \ \ \ \ in\ (case\ find\_aug\_path\ (quotG\ E)\ (quotG\ M)\ of\ Some\ p{\isacharprime}\ {\isasymRightarrow}\ Some\ (refine\ p{\isacharprime})\isanewline
\ \ \ \ \ \ \ \ \ \ \ \ \ \ \ {\isacharbar}\ \_\ {\isasymRightarrow}\ None)\isanewline
\ \ \ \ \ {\isacharbar}\ \_\ {\isasymRightarrow}\ None)\isanewline

\isacommand{lemma}\isamarkupfalse%
\ find\_aug\_path\_sound:\isanewline
\ \ \isakeyword{assumes}\isanewline
\ \ \ \ matching\ M\ \isakeyword{and} M\ {\isasymsubseteq}\ E\ \isakeyword{and} finite\ M\isanewline
\ \ \ \ \isakeyword{and}\isanewline
\ \ \ \ {\isasymforall}e{\isasymin}E.\ {\isasymexists}u\ v.\ e\ =\ \{u,\ v\}\ {\isasymand}\ u\ {\isasymnoteq}\ v\ \isakeyword{and}\ finite\ (Vs\ E)\isanewline
\ \ \ \ \isakeyword{and}\isanewline
\ \ \ \ find\_aug\_path\ E\ M\ =\ Some\ p\isanewline
\ \ \isakeyword{shows}\ augmenting\_path\ M\ p\ {\isasymand}\ path\ E\ p\ {\isasymand}\ distinct\ p\isanewline

\isacommand{lemma}\isamarkupfalse%
\ find\_aug\_path\_complete:\isanewline
\ \ \isakeyword{assumes}\isanewline
\ \ \ \ augmenting\_path\ M\ p\ \isakeyword{and} path\ E\ p\ \isakeyword{and} distinct\ p\isanewline
\ \ \ \ \isakeyword{and}\isanewline
\ \ \ \ matching\ M\ \isakeyword{and} M\ {\isasymsubseteq}\ E\ \isakeyword{and} finite\ M\isanewline
 \ \ \ \ \isakeyword{and}\isanewline
\ \ \ \ {\isasymforall}e{\isasymin}E.\ {\isasymexists}u\ v.\ e\ =\ \{u,\ v\}\ {\isasymand}\ u\ {\isasymnoteq}\ v\ \isakeyword{and} finite\ (Vs\ E)"\isanewline
\ \ \isakeyword{shows}\ {\isasymexists}p{\isacharprime}.\ find\_aug\_path\ E\ M\ =\ Some\ p{\isacharprime}

\end{isabelle}

Note that in \isa{find\_aug\_path}, we instantiate both arguments \isa{P} an \isa{s} of the locale \isa{quot} to obtain the quotienting function \isa{quotG} and the function for refining augmenting path \isa{refine}.

Lastly, what follows shows the validity of instantiating the functional argument of \isa{find\_max\_matching} with \isa{find\_aug\_path}, which gives us the following soundness theorem of the resulting algorithm.

\begin{isabelle}
\isacommand{lemma}\isamarkupfalse%
\ find\_max\_matching\_works:\isanewline
\ \ \isakeyword{assumes}\isanewline
\ \ \ \ finite\ (Vs\ E)\ \isakeyword{and} {\isasymforall}e{\isasymin}E.\ {\isasymexists}u\ v.\ e\ =\ \{u,\ v\}\ {\isasymand}\ u\ {\isasymnoteq}\ v\isanewline
\ \ \isakeyword{shows}\isanewline
\ \ \ \ find\_max\_match.find\_max\_matching\ find\_aug\_path\ E\ \{\}\ {\isasymsubseteq}\ E\isanewline
\ \ \ \ matching\ (find\_max\_match.find\_max\_matching\ find\_aug\_path\ E\ \{\})\isanewline
\ \ \ \ finite\ (find\_max\_match.find\_max\_matching\ find\_aug\_path\ E\ \{\})\isanewline
\ \ \ \ {\isasymforall}M.\ matching\ M\ {\isasymand}\ M\ {\isasymsubseteq}\ E\ {\isasymand}\ finite\ M\isanewline
\ \ \ \ \ \ \ \ {\isasymlongrightarrow}\ card\ M\ {\isasymle}\ card\ (find\_max\_match.find\_max\_matching\ find\_aug\_path\ E\ \{\})
\end{isabelle}

\subsection{Computing Blossoms and Augmenting Paths}

Until now, we have only assumed the existence of the function $\BlossomOrAugPath$, which can compute augmenting paths or blossoms, if any exist in the graph.
We now refine that to an algorithm which, given two alternating paths resulting from the ascent of alternating trees, returns either an augmenting path or a blossom.

\newcommand{\compref}{\textsf{\upshape longest\_disj\_pref}}
\newcommand{\drop}{\textsf{\upshape drop}}
\newcommand{\hd}{\textsf{\upshape first}}
\newcommand{\reverse}{\textsf{\upshape rev}}
We first introduce some notions and notation.
For a list $l$, let $\cardinality{l}$ be the length of $l$.
For a list $l$ and a natural number $n$, let $\drop\; n\; l$ denote the list $l$, but with the first $n$ elements dropped.
For a list $l$, let $h::l$ denote adding an element $h$ to the front of a list $l$.
For a non-empty list $l$, let $\hd\; l$  and $\last\; l$ denote the first and last elements of $l$, respectively.
Also, for a list $l$, let $\reverse\; l$ denote its reverse.
For two lists $l_1$ and $l_2$, let $l_1\cat l_2$ denote their concatenation.
Also, let $\compref\; l_1\; l_2$ denote the pair of lists $\langle l_1',l_2'\rangle$, where $l_1'$ and $l_2'$ are the longest disjoint prefixes of $l1$ and $l_2$, respectively, s.t.\ $\last\;l_1' = \last\;l_2'$.
Note: $\compref\; l_1\; l_2$ is only well-defined if there is are $l_1',l_2',$ and $l$ s.t.\ $l_1 = l_1'\cat l$ and $l_2 = l_2'\cat l$, and if both $l'_1$ and $l'_2$ are disjoint except at their endpoints.

\newcommand{\consblossom}{$\langle \reverse(\drop\;(\cardinality{\path_1'}-1)\;\path_1), (\reverse\; \path_1') \cat \path_2'\rangle$}

We now are able to state the following two lemmas concerning the construction of a blossom or an augmenting path given paths resulting from alternating trees search.

\begin{mylem}
\label{lem:consBlos}
If $\path_1$ and $\path_2$ are both (1) simple paths w.r.t. $\graph$, (2) alternating paths w.r.t. $\matching$, and (3) of odd length, and if we have that (4) $\last\;\path_1=\last\;\path_2$, (5)  $\last\;\path_1\not\in\bigcup\matching$, \item $\{\hd\;\path_1,\hd\;\path_2\}\in\graph$, (5) $\{\hd\;\path_1,\hd\;\path_2\}\not\in\matching$, and (7) $\compref\;\path_1\;\path_2$ is well-defined and  $\langle\path_1',\path_2'\rangle=\compref\;\path_1\;\path_2$, then \consblossom\\ is a blossom w.r.t. $\langle\graph,\matching\rangle$.
\end{mylem}

\begin{mylem}
\label{lem:consAugPath}
If $\path_1$ and $\path_2$ are both (1) simple paths w.r.t. $\graph$, (2) alternating paths w.r.t. $\matching$, (3)  of odd length, and (4) disjoint, and if we have that (5) $\last\;\path_1\not\in\bigcup\matching$, (6) $\last\;\path_2\not\in\bigcup\matching$, (7) $\last\;\path_1\neq\last\;\path_2$, (8) $\{\hd\;\path_1,\hd\;\path_2\}\in\graph$, and (9)  $\{\hd\;\path_1,\hd\;\path_2\}\not\in\matching$, then $(\reverse\;\path_1)\cat\path_2$ is an augmenting path w.r.t. $\langle\graph,\matching\rangle$.
\end{mylem}

\noindent Based on the above lemmas we refine the algorithm $\BlossomOrAugPath$ as shown in Algorithm~\ref{alg:BlossomOrAug}.

\newcommand{\computaltpath}{\textsf{\upshape compute\_alt\_path}}
\renewcommand{\choice}{\textsf{\upshape choose}\;}

\SetKwIF{If}{ElseIf}{Else}{if}{}{else if}{else}{endif}
\begin{algorithm}\DontPrintSemicolon
  \KwSty{if} $\exists e\in\graph. e \cap \bigcup \matching = \emptyset$\;
  \ \ \KwSty{return} Augmenting path $\choice \{e\mid e\in\graph \wedge e \cap \bigcup \matching = \emptyset\}$\;
  \KwSty{else} \KwSty{if} $\computaltpath(\graph,\matching) = \langle\path_1,\path_2\rangle$\;
  \ \ \KwSty{if} $\last\;\path_1\neq\last\;\path_2$\;
  \ \ \ \ \KwSty{return} Augmenting path $(\reverse\;\path_1)\cat\path_2$\;
  \ \ \KwSty{else}\;
  \ \ \ \ $\langle\path_1',\path_2'\rangle=\compref\;\path_1\;\path_2$\;
  \ \ \ \ \KwSty{return} Blossom \consblossom\;
  \KwSty{else}\;
  \ \ \KwSty{return} No blossom or augmenting path found
  \caption{$\BlossomOrAugPath(\graph, \matching)$}\label{alg:BlossomOrAug}
\end{algorithm}

The following corollary shows the conditions under which $\BlossomOrAugPath$ works.

\begin{mycor}
\label{cor:computaltpath}
Assume the function $\computaltpath(\graph,\matching)$ returns two lists of vertices $\langle\path_1,\path_2\rangle$ s.t.\ both lists are (1) simple paths w.r.t.\ $\graph$, (2) alternating paths w.r.t.\ $\matching$, and (3) of odd length, and also (4) $\last\;\path_1\not\in\bigcup\matching$, (5) $\last\;\path_2\not\in\bigcup\matching$, (6) $\{\hd\;\path_1,\hd\;\path_2\}\in\graph$, and (7) $\{\hd\;\path_1,\hd\;\path_2\}\not\in\matching$, iff two lists of vertices with those properties exist.
Then there is a blossom or an augmenting path w.r.t.\ $\langle\graph,\matching\rangle$ iff $\BlossomOrAugPath(\graph, \matching)$ is a blossom or an augmenting path w.r.t.\ $\langle\graph,\matching\rangle$.
\end{mycor}
In Isabelle/HOL, to formalise the function $\BlossomOrAugPath$, we firstly defined a function, \isa{longest\_disj\_pfx}, which finds the longest common prefix in a straightforward fashion with a quadratic wort-case runtime.
The formalised versions of Lemma~\ref{lem:consBlos}~and~\ref{lem:consAugPath}, which show that the output of \isa{longest\_disj\_pfx} can be used to construct a blossom or an augmenting path are as follows:

\begin{isabelle}
\isacommand{lemma}\isamarkupfalse%
\ common\_pfxs\_form\_blossom:\isanewline
\ \ \isakeyword{assumes}\isanewline
\ \ \ \ (Some\ pfx{\isadigit{1}},\ Some\ pfx{\isadigit{2}})\ =\ longest\_disj\_pfx\ p{\isadigit{1}}\ p{\isadigit{2}}"\isanewline
\ \ \ \ \isakeyword{and}\isanewline
\ \ \ \ p{\isadigit{1}}\ =\ pfx{\isadigit{1}}\ {\isacharat}\ p\ \isakeyword{and} p{\isadigit{2}}\ =\ pfx{\isadigit{2}}\ {\isacharat}\ p"\isanewline
\ \ \ \ \isakeyword{and}\isanewline
\ \ \ \ alt\_path\ M\ p{\isadigit{1}}\ \isakeyword{and} alt\_path\ M\ p{\isadigit{2}}\ \isakeyword{and} last\ p{\isadigit{1}}\ {\isasymnotin}\ Vs\ M\ \isakeyword{and} \{hd\ p{\isadigit{1}},\ hd\ p{\isadigit{2}}\}\ {\isasymin}\ M"\isanewline
\ \ \ \ \isakeyword{and}\ \isanewline
\ \ \ \ hd\ p{\isadigit{1}}\ {\isasymnoteq}\ hd\ p{\isadigit{2}}"\isanewline
\ \ \ \ \isakeyword{and}\isanewline
\ \ \ \ even\ (length\ p{\isadigit{1}})\ \isakeyword{and} even\ (length\ p{\isadigit{2}})\isanewline
\ \ \ \ \isakeyword{and}\isanewline
\ \ \ \ distinct\ p{\isadigit{1}}\ \isakeyword{and} distinct\ p{\isadigit{2}}\isanewline
\ \ \ \ \isakeyword{and}\isanewline
\ \ \ \ matching\ M\isanewline
\ \ \isakeyword{shows}\ blossom\ M\ (rev\ (drop\ (length\ pfx{\isadigit{1}})\ p{\isadigit{1}}))\ (rev\ pfx{\isadigit{1}}\ {\isacharat}\ pfx{\isadigit{2}})\isanewline

\isacommand{lemma}\isamarkupfalse%
\ construct\_aug\_path:\isanewline
\ \ \isakeyword{assumes}\isanewline
\ \ \ \ set\ p{\isadigit{1}}\ {\isasyminter}\ set\ p{\isadigit{2}}\ =\ \{\}\isanewline
\ \ \ \ \isakeyword{and}\isanewline
\ \ \ \ p{\isadigit{1}}\ {\isasymnoteq}\ {\isacharbrackleft}{\isacharbrackright}\ \isakeyword{and} p{\isadigit{2}}\ {\isasymnoteq}\ {\isacharbrackleft}{\isacharbrackright}\isanewline
\ \ \ \ \isakeyword{and}\isanewline
\ \ \ \ alt\_path\ M\ p{\isadigit{1}}\ \isakeyword{and} alt\_path\ M\ p{\isadigit{2}}\ \isakeyword{and} last\ p{\isadigit{1}}\ {\isasymnotin}\ Vs\ M\ \isakeyword{and} last\ p{\isadigit{2}}\ {\isasymnotin}\ Vs\ M\isanewline
\ \ \ \ \isakeyword{and}\isanewline
\ \ \ \ \{hd\ p{\isadigit{1}},\ hd\ p{\isadigit{2}}\}\ {\isasymin}\ M\ \isakeyword{and}\isanewline
\ \ \ \ \isakeyword{and}\isanewline
\ \ \ \ even\ (length\ p{\isadigit{1}})\ \isakeyword{and} even\ (length\ p{\isadigit{2}})\isanewline
\ \ \isakeyword{shows}\ augmenting\_path\ M\ ((rev\ p{\isadigit{1}})\ {\isacharat}\ p{\isadigit{2}})
\end{isabelle}

The function $\BlossomOrAugPath$ is formalised as follows:
\begin{isabelle}
\ \ "compute\_blossom\ G\ M\ {\isasymequiv}\ \isanewline
\ \ \ (if\ ({\isasymexists}e.\ e\ {\isasymin}\ unmatched\_edges\ G\ M)\ then\isanewline
\ \ \ \ \ \ \ \ \ let\isanewline
\ \ \ \ \ \ \ \ \ \ \ singleton\_path\ =\isanewline
\ \ \ \ \ \ \ \ \ \ \ \ \ \ (SOME\ p.\ {\isasymexists}v{\isadigit{1}}\ v{\isadigit{2}}.\ p\ =\ {\isacharbrackleft}v{\isadigit{1}}\ ,v{\isadigit{2}}{\isacharbrackright}\ {\isasymand}\ \{v{\isadigit{1}},\ v{\isadigit{2}}\}\ {\isasymin}\ unmatched\_edges\ G\ M)\isanewline
\ \ \ \ \ \ \ \ \ in\isanewline
\ \ \ \ \ \ \ \ \ \ \ Some\ (Path\ singleton\_path)\isanewline
\ \ \ \ else\isanewline
\ \ \ \ \ case\ compute\_alt\_path\ G\ M\isanewline
\ \ \ \ \ \ \ of\ Some\ (p{\isadigit{1}},p{\isadigit{2}})\ {\isasymRightarrow}\ \isanewline
\ \ \ \ \ \ \ \ \ (if\ (set\ p{\isadigit{1}}\ {\isasyminter}\ set\ p{\isadigit{2}}\ =\ \{\})\ then\isanewline
\ \ \ \ \ \ \ \ \ \ \ \ Some\ (Path\ ((rev\ p{\isadigit{1}})\ {\isacharat}\ p{\isadigit{2}}))\isanewline
\ \ \ \ \ \ \ \ \ \ else\isanewline
\ \ \ \ \ \ \ \ \ \ \ \ (let\isanewline
\ \ \ \ \ \ \ \ \ \ \ \ \ \ \ (pfx{\isadigit{1}},\ pfx{\isadigit{2}})\ =\ longest\_disj\_pfx\ p{\isadigit{1}}\ p{\isadigit{2}}{\isacharsemicolon}\isanewline
\ \ \ \ \ \ \ \ \ \ \ \ \ \ \ stem\ =\ (rev\ (drop\ (length\ (the\ pfx{\isadigit{1}}))\ p{\isadigit{1}})){\isacharsemicolon}\isanewline
\ \ \ \ \ \ \ \ \ \ \ \ \ \ \ cycle\ =\ (rev\ (the\ pfx{\isadigit{1}})\ {\isacharat}\ (the\ pfx{\isadigit{2}}))\isanewline
\ \ \ \ \ \ \ \ \ \ \ \ \ in\isanewline
\ \ \ \ \ \ \ \ \ \ \ \ \ \ (Some\ (Blossom\ stem\ cycle))))\isanewline
\ \ \ \ \ \ \ {\isacharbar}\ \_\ {\isasymRightarrow}\ None)"
\end{isabelle}

We use a locale again to formalise that function.
That locale parameterises it on a function that searches for alternating paths and poses the soundness and completeness assumptions for that alternating path search function.
This function is equivalent to the unspecified function $\computaltpath$ in Corollary~\ref{cor:computaltpath} and locale's assumptions on it are formalised statements of the seven assumptions on $\computaltpath$ in Corollary~\ref{cor:computaltpath}.

\subsection{Computing Alternating Paths}
{
\newcommand{\lab}{\textsf{\upshape label}}
\newcommand{\even}{\textsf{\upshape even}}
\newcommand{\odd}{\textsf{\upshape odd}}
\newcommand{\examined}{\textsf{\upshape ex}}
\newcommand{\unexamined}{\textsf{\upshape unex\_with\_even}}
\newcommand{\parent}{\textsf{\upshape parent}}
\newcommand{\follow}[1]{\textsf{\upshape follow}\;#1\;}
\renewcommand{\comment}[1]{{\scriptsize // #1}}

Lastly, we refine the function $\computaltpath$ to an algorithmic specification.
The algorithmic specification of that function performs the alternating tree search, see Algorithm~\ref{alg:compAltPath}.
If the function positively terminates, i.e. finding two vertices with even labels, returns two alternating paths by ascending the two alternating trees to which the two vertices belong.
This tree ascent is performed by the function $\follow{}$.
That function takes a functional argument $f$ and a vertex, and returns the singleton list $[\vertexgen]$ if $f(\vertexgen)=$ None, and $\vertexgen::(\follow{f}(f(\vertexgen)))$ otherwise.

\SetKwIF{If}{ElseIf}{Else}{if}{}{else if}{else}{endif}
\begin{algorithm}\DontPrintSemicolon
  $\examined=\emptyset$ \comment{Set of examined edges}\;
  \KwSty{foreach} $\vertexgen\in \bigcup \graph$\;
  \ \ \ \ $\lab\;\vertexgen = $ None\;
  \ \ \ \ $\parent\;\vertexgen = $ None\;
  $U = \bigcup\graph \setminus\bigcup \matching$ \comment{Set of unmatched vertices}\;
  \KwSty{foreach} $\vertexgen\in U$\;
  \ \ \ \ $\lab\;\vertexgen = \langle u,\even\rangle$\;
  \KwSty{while} $(\graph\setminus \examined) \cap \{e\mid\exists \vertexgen\in e, r\in\bigcup\graph. \lab\; \vertexgen = \langle r, \even\rangle\}\neq\emptyset$\;
  \ \ \ \ \comment{Choose a new edge and labelled it examined}\;
  \ \ \ \ $\{\vertexa,\vertexb\}=\choice (\graph\setminus \examined) \cap \{\{\vertexa,\vertexb\}\mid\exists r. \lab\; \vertexa = \langle r, \even\rangle\}$\;
  \ \ \ \ $\examined$ = $\examined \cup \{\{\vertexa,\vertexb\}\}$\;
  \ \ \ \ \KwSty{if} $\lab\; \vertexb = $ None\;
  \ \ \ \ \ \ \ \ \comment{Grow the discovered set of edges from $r$ by two}\;
  \ \ \ \ \ \ \ \ $\vertexc=\choice \{\vertexc \mid \{\vertexb, \vertexc\} \in \matching\}$\;
  \ \ \ \ \ \ \ \ $\examined$ = $\examined \cup \{\{\vertexb,\vertexc\}\}$\;
  \ \ \ \ \ \ \ \ $\lab\; \vertexb = \langle r, \odd \rangle$; $\lab\; \vertexc = \langle r, \even\rangle$; $\parent\;\vertexb = \vertexa$; $\parent\;\vertexc = \vertexb$\;
  \ \ \ \ \KwSty{else} \KwSty{if} $\exists s\in\bigcup\graph. \lab\; \vertexb = \langle s, \even\rangle$\;
  \ \ \ \ \ \ \ \ \comment{Return two paths from current edge's tips to unmatched vertex(es)}\;
  \ \ \ \ \ \ \ \ \KwSty{return} $\langle\follow{\parent} \vertexa, \follow{\parent} \vertexb\rangle$\;
  \KwSty{return} No paths found\;
  \caption{$\computaltpath(\graph,\matching)$}\label{alg:compAltPath}
\end{algorithm}
The soundness and completeness of Algorithm~\ref{alg:compAltPath}  is stated as follows.
\begin{mythm}
The function $\computaltpath(\graph,\matching)$ returns two lists of vertices $\langle\path_1,\path_2\rangle$ s.t.\ both lists are (1) simple paths w.r.t. $\graph$, (2) alternating paths w.r.t. $\matching$, and (3) of odd length, and also (4) $\last\;\path_1\not\in\bigcup\matching$, (5) $\last\;\path_2\not\in\bigcup\matching$, (6) $\{\hd\;\path_1,\hd\;\path_2\}\in\graph$, and (7) $\{\hd\;\path_1,\hd\;\path_2\}\not\in\matching$, iff two lists of vertices with those properties exist.
\end{mythm}

The primary difficulty with proving this theorem is identifying the loop invariants, which are as follows:
\begin{enumerate}
  \item For any vertex $\vertexgen$, if for some $r$, $\lab\;\vertexgen = \langle r, \even\rangle$, then the vertices in the list $\follow{\parent}\vertexgen$ have labels that alternate between $\langle r, \even\rangle$ and $\langle r, \odd\rangle$.
  \item For any vertex $\vertexa$, if for some $r$ and some $l$, we have $\lab\;\vertexa=\langle r,l\rangle$, then the vertex list $\vertexa\vertexb\dots\vertexgen_n$ returned by $\follow{\parent} \vertexa$ has the following property: if $\lab\;\vertexgen_i = \langle r, \even\rangle$ and $\lab\;\vertexgen_{i+1} = \langle r, \odd\rangle$, for some $r$, then $\{\vertexgen_i,\vertexgen_{i+1}\}\in\matching$, otherwise, $\{\vertexgen_i,\vertexgen_{i+1}\}\not\in\matching$.
  \item The relation induced by the function $\parent$ is well-founded.
  \item For any $\{\vertexa,\vertexb\}\in\matching$, $\lab\;\vertexa=$ None iff $\lab\;\vertexb=$ None.
  \item For any $\vertexa$, if $\lab\;\vertexa=$ None then $\parent\;\vertexb\neq\vertexa$, for all $\vertexb$.
  \item For any $\vertexgen$, if $\lab\;\vertexgen\neq$ None, then $\last\;(\follow{\parent} \vertexgen)\not\in\bigcup\matching$.

  \item For any $\vertexgen$, if $\lab\;\vertexgen\neq$ None, then $\lab\;(\last\;(\follow{\parent} \vertexgen))=\langle r, \even\rangle$, for some $r$.

  \item For any $\{\vertexa,\vertexb\}\in\matching$, if $\lab\; \vertexa\neq$ None, then $\{\vertexa,\vertexb\}\in\examined$.
  \item For any $\vertexgen$, $\follow{\parent} \vertexgen$ is a simple path w.r.t.\ $\graph$.
  \item Suppose we have two vertex lists $\path_1$ and $\path_2$, s.t. both lists are (1)  simple paths w.r.t.\ $\graph$, (2) alternating paths w.r.t.\ $\matching$, and (3) of odd length, and also (4) $\last\;\path_1\not\in\bigcup\matching$, (5) $\last\;\path_2\not\in\bigcup\matching$, (6) $\{\hd\;\path_1,\hd\;\path_2\}\in\graph$, and (7) $\{\hd\;\path_1,\hd\;\path_2\}\not\in\matching$.
Then there is at least an edge from the path $\reverse\;\path_1\cat\path_2$ which is a member of neither $\matching$ nor $\examined$.%
\footnote{The hypothesis of this invariant is equivalent to the existence of an augmenting path or a blossom w.r.t. $\langle \graph,\matching\rangle$.}
\end{enumerate}

To formalise Algorithm~\ref{alg:compAltPath} in Isabelle/HOL, we first define the function which follows a vertex's parent as follows:
\begin{isabelle}
\ \ follow\ v\ =\ (case\ (parent\ v)\ of\ Some\ v{\isacharprime}\ {\isasymRightarrow}\ v\ {\isacharhash}\ (follow\ v{\isacharprime})\ {\isacharbar}\ \_\ \ {\isasymRightarrow}\ {\isacharbrackleft}v{\isacharbrackright})
\end{isabelle}
Again, we use a locale to formalise that function, and that locale fixes the function \isa{parent}.
Note that the above function is not well-defined for all possible arguments.
In particular, it is only well-defined if the relation between pairs of vertices induced by the function \isa{parent} is a well-founded relation.
This assumption on \isa{parent} is a part of the locale's definition.

Then, we then formalise $\computaltpath$ as follows:
\begin{isabelle}
\ compute\_alt\_path\ ex\ par\ flabel\ =\ \isanewline
\ \ \ \ (if\ ({\isasymexists}v{\isadigit{1}}\ v{\isadigit{2}}.\ \{v{\isadigit{1}},\ v{\isadigit{2}}\}\ {\isasymin}\ G\ -\ ex\ {\isasymand}\ ({\isasymexists}r.\ flabel\ v{\isadigit{1}}\ =\ Some\ (r,\ Even)))\ then\isanewline
\ \ \ \ \ \ \ let\isanewline
\ \ \ \ \ \ \ \ \ (v{\isadigit{1}},v{\isadigit{2}})\ =\ (SOME\ (v{\isadigit{1}},v{\isadigit{2}}).\ \{v{\isadigit{1}},\ v{\isadigit{2}}\}\ {\isasymin}\ G\ -\ ex\ {\isasymand}\isanewline
\ \ \ \ \ \ \ \ \ \ \ \ \ \ \ \ \ \ \ \ \ \ \ \ \ \ \ \ \ \ \ \ \ \ ({\isasymexists}r.\ flabel\ v{\isadigit{1}}\ =\ Some\ (r,\ Even))){\isacharsemicolon}\isanewline
\ \ \ \ \ \ \ \ \ ex{\isacharprime}\ =\ insert\ \{v{\isadigit{1}},\ v{\isadigit{2}}\}\ ex{\isacharsemicolon}\isanewline
\ \ \ \ \ \ \ \ \ r\ =\ (SOME\ r.\ flabel\ v{\isadigit{1}}\ =\ Some\ (r,\ Even))\isanewline
\ \ \ \ \ \ \ in\isanewline
\ \ \ \ \ \ \ \ \ (if\ flabel\ v{\isadigit{2}}\ =\ None\ {\isasymand}\ ({\isasymexists}v{\isadigit{3}}.\ \{v{\isadigit{2}},\ v{\isadigit{3}}\}\ {\isasymin}\ M)\ then\isanewline
\ \ \ \ \ \ \ \ \ \ \ let\isanewline
\ \ \ \ \ \ \ \ \ \ \ \ \ v{\isadigit{3}}\ =\ (SOME\ v{\isadigit{3}}.\ \{v{\isadigit{2}},\ v{\isadigit{3}}\}\ {\isasymin}\ M){\isacharsemicolon}\isanewline
\ \ \ \ \ \ \ \ \ \ \ \ \ par{\isacharprime}\ =\ par(v{\isadigit{2}}\ :=\ Some\ v{\isadigit{1}},\ v{\isadigit{3}}\ :=\ Some\ v{\isadigit{2}}){\isacharsemicolon}\isanewline
\ \ \ \ \ \ \ \ \ \ \ \ \ flabel{\isacharprime}\ =\ flabel(v{\isadigit{2}}\ :=\ Some\ (r,\ Odd),\ v{\isadigit{3}}\ :=\ Some\ (r,\ Even)){\isacharsemicolon}\isanewline
\ \ \ \ \ \ \ \ \ \ \ \ \ ex{\isacharprime}{\isacharprime}\ =\ insert\ \{v{\isadigit{2}},\ v{\isadigit{3}}\}\ ex{\isacharprime}{\isacharsemicolon}\isanewline
\ \ \ \ \ \ \ \ \ \ \ \ \ return\ =\ compute\_alt\_path\ ex{\isacharprime}{\isacharprime}\ par{\isacharprime}\ flabel{\isacharprime}\isanewline
\ \ \ \ \ \ \ \ \ \ \ in\isanewline
\ \ \ \ \ \ \ \ \ \ \ \ \ return\isanewline
\ \ \ \ \ \ \ \ \ else\ if\ {\isasymexists}r.\ flabel\ v{\isadigit{2}}\ =\ Some\ (r,\ Even)\ then\isanewline
\ \ \ \ \ \ \ \ \ \ \ \ let\isanewline
\ \ \ \ \ \ \ \ \ \ \ \ \ \ r{\isacharprime}\ =\ (SOME\ r{\isacharprime}.\ flabel\ v{\isadigit{2}}\ =\ Some\ (r{\isacharprime},\ Even)){\isacharsemicolon}\ \isanewline
\ \ \ \ \ \ \ \ \ \ \ \ \ \ return\ =\ Some\ (parent.follow\ par\ v{\isadigit{1}},\ parent.follow\ par\ v{\isadigit{2}})\isanewline
\ \ \ \ \ \ \ \ \ \ \ \ in\isanewline
\ \ \ \ \ \ \ \ \ \ \ \ \ \ return\isanewline
\ \ \ \ \ \ \ \ \ else\isanewline
\ \ \ \ \ \ \ \ \ \ \ let\isanewline
\ \ \ \ \ \ \ \ \ \ \ \ \ \ return\ =\ None\isanewline
\ \ \ \ \ \ \ \ \ \ \ in\isanewline
\ \ \ \ \ \ \ \ \ \ \ \ \ return)\isanewline
\ \ \ \ \ else\isanewline
\ \ \ \ \ \ \ let\isanewline
\ \ \ \ \ \ \ \ \ return\ =\ None\isanewline
\ \ \ \ \ \ \ in\isanewline
\ \ \ \ \ \ \ \ \ return)
\end{isabelle}

Note that we do not use a while combinator to represent the while loop: instead we formalise it recursively, passing the context along recursive calls.
In particular, we define it as a recursive function which takes as arguments the variables representing the state of the while loop, namely, the set of examined edges \isa{ex}, the parent function \isa{par}, and the labelling function \isa{flabel}.


    
}

\subsection{Discussion}
The algorithm in LEDA differs from the description above in one aspect. If no augmenting path is found, an odd-set cover is constructed proving optimality. Also the correctness proof uses the odd-set cover instead of the fact that an augmenting path exists in the original graph if and only if one exists in the quotient graph.

For an efficient implementation, the shrinking process and the lifting of augmenting paths are essential. The shrinking process is implemented using a union-find data structure and the lifting is supported by having each node in a contracted cycle point to the edge that closes the cycle in a blossom~\cite{LEDAbook}.

\section{Level Five of Trustworthiness: Extraction of Efficient Executable Code}

In this section we examine the process of obtaining trustworthy executable
and efficient code from algorithms verified in theorem provers. First we
discuss the problem in general and then we examine our formalization of the
blossom-shrinking algorithm.

Most theorem provers are connected to a programming language of some sort.
Frequently, as in the case of Isabelle/HOL, that programming language is a
subset of the logic and close to a functional programming language.  The
theorem prover will usually support the extraction of actual code in some
programming language. Isabelle/HOL supports Standard ML, Haskell,
OCaml and Scala.

To show that code extraction ``works'', here are some random non-trivial examples of verifications that have resulted
in reasonably efficient code: Compilers for C \cite{Leroy-JAR09}
and for ML~\cite{KumarMNO14}, a SPIN-like
model checker~\cite{cav/EsparzaLNNSS13}, network flow
algorithms~\cite{jar/LammichS19} and the Berlekamp-Zassenhaus factorization
algorithm~\cite{jar/DivasonJTY19}.

We will now discuss some approaches to obtaining code from function
definitions in a theorem prover. In the ACL2 theorem prover all functions are
defined in a purely functional subset of Lisp and are thus directly
executable. In other systems, code generation involves an explicit
translation step. The trustworthiness of this step varies.  Probably the most
trustworthy code generator is that of HOL4, because its backend is a verified
compiler for CakeML~\cite{KumarMNO14}, a dialect of ML. The step from HOL to
CakeML is not verified once and for all, but every time it is run it produces
a theorem that can be examined and that states the correctness of this
run~\cite{jfp/MyreenO14}.
The standard code generator in Isabelle/HOL is unverified (although its
underlying theory has been proved correct on
paper~\cite{HaftmannN-FLOPS2010}).  There is ongoing work to replace it with
a verified code generator that produces CakeML~\cite{HupelN-ESOP18}.

So far we have only considered purely functional code but efficient
algorithms often make use of imperative features. Some theorem provers
support imperative languages directly, e.g. Java~\cite{lncs/10001}.  We will
now discuss how to generate imperative code from purely functional
one. Clearly the code generator must turn the purely functional definitions
into more imperative ones. The standard approach~\cite{BulwahnKHEM08,jfp/MyreenO14} is to let the code generator
recognize monadic definitions (a purely functional way to express imperative computations) and implement those imperatively. This is possible because
many functional programming languages do in fact offer imperative features as well.

Just as important as the support for code extraction is the support for
verified stepwise refinement of data types and algorithms by the user.
Data refinement means the replacement of abstract data types by
concrete efficient ones, e.g.\ sets by search trees. Algorithm refinement
means the stepwise replacement of abstract high-level definitions that may
not even be executable by efficient implementations.
Both forms of refinement are supported well in
Isabelle/HOL~\cite{HaftmannKKN-ITP13,Lammich13,jar/Lammich19}.

We conclude this section with a look at code generation from our formalization
of the blossom-shrinking algorithm. It turns out that our formalization is almost
executable as is.  The only non-executable construct we used is
\isa{SOME\ x.\;P} that denotes some arbitrary \isa{x} that satisfies the
predicate \isa{P}. Of course one can hide arbitrarily complicated
computations in such a contruct but we have used it only for simple
nondeterministic choices and it will be easy to replace.
For example, one can obtain an executable version of function
\isa{choose\_con\_vert} (see Section~\ref{subsec:Contraction}) by defining a
function that searches the vertex list \isa{vs} for the first
\isa{v'} such that \isa{{\isacharbraceleft}v{\isacharcomma}\ v{\isacharprime}{\isacharbraceright}\ {\isasymin}\ E}. This is an example of algorithm refinement.
To arrive at efficient code for the blossom-shrinking algorithm as a whole we will need to
apply both data and algorithm refinement down to the imperative level. At
least the efficient implementations referred to above, just before
Section~\ref{sec:Isabelle}, are intrinsically imperative.

Finally let us note that instead of code generation it is also possible to
verify existing code in a theorem prover. This was briefly mentioned in
Section~\ref{sec:level3} and Chargu{\'{e}}raud~\cite{icfp/Chargueraud11}
has followed this approach quite successfully.

}

\section{The Future}

The state of the art in the verification of complex algorithms has improved
enormously over the last decade. Yet there is still a lot to do on the path to
a verified library such as LEDA. Apart from the shere amount of material that
would have to be verified there is the challenge of obtaining
trustworthy code that is of comparable efficiency. This requires trustworthy
code generation for a language such  C or C++, including the memory management.
This is a non-trivial task, but some of the pieces of the puzzle,
like a verified compiler, are in place already.


\bibliographystyle{plainurl}
\bibliography{ref,MFCS,references}

\begin{thebibliography}{10}

\bibitem{BlossomCorrectnessProof}
Mohammad Abdulaziz, Kurt Mehlhorn, and Tobias Nipkow.
\newblock A correctness proof of {E}dmonds' blossom shrinking algorithm for
  maximum matchings in graphs.
\newblock forthcoming.

\bibitem{DBLP:journals/jar/AbdulazizNG18}
Mohammad Abdulaziz, Michael Norrish, and Charles Gretton.
\newblock Formally verified algorithms for upper-bounding state space
  diameters.
\newblock {\em J. Autom. Reasoning}, 61(1-4):485--520, 2018.

\bibitem{DBLP:journals/jar/AbdulazizPaulson18}
Mohammad Abdulaziz and Lawrence Paulson.
\newblock An {I}sabelle/{HOL} formalisation of {G}reen's theorem.
\newblock {\em J. Autom. Reasoning, In Press}.

\bibitem{lncs/10001}
Wolfgang Ahrendt, Bernhard Beckert, Richard Bubel, Reiner H{\"{a}}hnle,
  Peter~H. Schmitt, and Mattias Ulbrich, editors.
\newblock {\em Deductive Software Verification - The KeY Book - From Theory to
  Practice}, volume 10001 of {\em Lecture Notes in Computer Science}.
\newblock Springer, 2016.

\bibitem{FrameworkVerificationCertifyingComputations}
E.~Alkassar, S.~B\"ohme, K.~Mehlhorn, and Ch. Rizkallah.
\newblock \htmladdnormallink{A Framework for the Verification of Certifying
  Computations}{http://arxiv.org/abs/1301.7462}.
\newblock {\em Journal of Automated Reasoning (JAR)}, 52(3):241--273, 2014.

\bibitem{Althaus-Mehlhorn}
E.~Althaus and K.~Mehlhorn.
\newblock \htmladdnormallink{Maximum Network Flow with Floating Point
  Arithmetic}{http://www.mpi-sb.mpg.de/\~{}mehlhorn/ftp/Althaus-Mehlhorn-maxflow.ps}.
\newblock {\em Information Processing Letters}, 66:109--113, 1998.

\bibitem{BulwahnKHEM08}
Lukas Bulwahn, Alexander Krauss, Florian Haftmann, Levent Erk{\"{o}}k, and John
  Matthews.
\newblock Imperative functional programming with {Isabelle/HOL}.
\newblock In {\em Theorem Proving in Higher Order Logics, 21st International
  Conference, TPHOLs 2008}, pages 134--149, 2008.

\bibitem{icfp/Chargueraud11}
Arthur Chargu{\'{e}}raud.
\newblock Characteristic formulae for the verification of imperative programs.
\newblock In {\em Proceeding of the 16th {ACM} {SIGPLAN} international
  conference on Functional Programming, {ICFP} 2011, Tokyo, Japan, September
  19-21, 2011}, pages 418--430, 2011.

\bibitem{Cohen:TPHOLs2009-23}
Ernie Cohen, Markus Dahlweid, Mark Hillebrand, Dirk Leinenbach, Micha{\l}
  Moskal, Thomas Santen, Wolfram Schulte, and Stephan Tobies.
\newblock {VCC}: {A} practical system for verifying concurrent {C}.
\newblock In {\em Theorem Proving in Higher Order Logics (TPHOLs 2009)}, volume
  5674 of {\em LNCS}, pages 23--42. Springer, 2009.

\bibitem{dahmen2019formalizing}
Sander~R Dahmen, Johannes H{\"o}lzl, and Robert~Y Lewis.
\newblock {Formalizing the Solution to the Cap Set Problem}.
\newblock {\em arXiv:1907.01449}, 2019.

\bibitem{jar/DivasonJTY19}
Jose Divas\'on, Sebastiaan J.~C. Joosten, Ren\'e Thiemann, and Akihisa Yamada.
\newblock A verified implementation of the {Berlekamp-Zassenhaus} factorization
  algorithm.
\newblock {\em J. Autom. Reasoning}, 2019.
\newblock published online.

\bibitem{Edmonds:matching}
J.~Edmonds.
\newblock Maximum matching and a polyhedron with 0,1 - vertices.
\newblock {\em Journal of Research of the National Bureau of Standards},
  69B:125--130, 1965.

\bibitem{cav/EsparzaLNNSS13}
Javier Esparza, Peter Lammich, Ren{\'{e}} Neumann, Tobias Nipkow, Alexander
  Schimpf, and Jan{-}Georg Smaus.
\newblock A fully verified executable {LTL} model checker.
\newblock In {\em Computer Aided Verification - 25th International Conference,
  {CAV} 2013}, pages 463--478, 2013.

\bibitem{Fortune96}
S.~Fortune.
\newblock Robustness issues in geometric algorithms.
\newblock In {\em Proceedings of the 1st Workshop on Applied Computational
  Geometry: Towards Geometric Engineering ({WACG}'96)}, volume 1148 of {\em
  {\rm Lecture Notes in Computer Science}}, pages 9--13, 1996.

\bibitem{Gabow:edmonds}
H.~N. Gabow.
\newblock An efficient implementation of {E}dmonds' algorithm for maximum
  matching on graphs.
\newblock {\em Journal of the ACM}, 23:221--234, 1976.

\bibitem{greenaway-etal:2012:gap}
David Greenaway, June Andronick, and Gerwin Klein.
\newblock Bridging the gap: Automatic verified abstraction of {C}.
\newblock In {\em Interactive Theorem Proving}, volume 7406 of {\em LNCS},
  pages 99--115, 2012.

\bibitem{HaftmannKKN-ITP13}
Florian Haftmann, Alexander Krauss, Ond\v{r}ej Kun\v{c}ar, and Tobias Nipkow.
\newblock Data refinement in {Isabelle/HOL}.
\newblock In S.~Blazy, C.~Paulin-Mohring, and D.~Pichardie, editors, {\em
  Interactive Theorem Proving (ITP 2013)}, volume 7998 of {\em LNCS Series},
  pages 100--115, 2013.

\bibitem{HaftmannN-FLOPS2010}
Florian Haftmann and Tobias Nipkow.
\newblock Code generation via higher-order rewrite systems.
\newblock In M.~Blume, N.~Kobayashi, and G.~Vidal, editors, {\em Functional and
  Logic Programming (FLOPS 2010)}, volume 6009 of {\em LNCS}, pages 103--117.
  Springer, 2010.

\bibitem{HupelN-ESOP18}
Lars Hupel and Tobias Nipkow.
\newblock A verified compiler from {Isabelle/HOL} to {CakeML}.
\newblock In A.~Ahmed, editor, {\em European Symposium on Programming (ESOP
  2018)}, volume 10801 of {\em LNCS}, pages 999--1026. Springer, 2018.

\bibitem{ClassRoomExample}
L.~Kettner, K.~Mehlhorn, S.~Pion, S.~Schirra, and C.~Yap.
\newblock \htmladdnormallink{Classroom Examples of Robustness Problems in
  Geometric
  Computations}{http://www.mpi-sb.mpg.de/\~{}mehlhorn/ftp/ClassRoomExample.ps}.
\newblock {\em Computational Geometry: Theory and Applications (CGTA)},
  40:61--78, 2008.
\newblock a preliminary version appeared in ESA 2004, LNCS 3221, pages 702 --
  713.

\bibitem{klein-etal:2010:sel4}
Gerwin Klein, June Andronick, Kevin Elphinstone, Gernot Heiser, David Cock,
  Philip Derrin, Dhammika Elkaduwe, Kai Engelhardt, Rafal Kolanski, Michael
  Norrish, Thomas Sewell, Harvey Tuch, and Simon Winwood.
\newblock {seL4}: Formal verification of an operating-system kernel.
\newblock {\em CACM}, 53(6):107--115, 2010.

\bibitem{Korte-Vygen}
B.~Korte and J.Vygen.
\newblock {\em Combinatorial Optimization: Theory and Algorithms}.
\newblock Springer, 2012.

\bibitem{KumarMNO14}
Ramana Kumar, Magnus~O. Myreen, Michael Norrish, and Scott Owens.
\newblock {CakeML}: a verified implementation of {ML}.
\newblock In {\em The 41st Annual {ACM} {SIGPLAN-SIGACT} Symposium on
  Principles of Programming Languages, {POPL} '14}, pages 179--192. {ACM},
  2014.

\bibitem{Lammich13}
Peter Lammich.
\newblock Automatic data refinement.
\newblock In {\em Interactive Theorem Proving - 4th International Conference,
  {ITP} 2013}, pages 84--99, 2013.

\bibitem{jar/Lammich19}
Peter Lammich.
\newblock Refinement to imperative {HOL}.
\newblock {\em J. Autom. Reasoning}, 62(4):481--503, 2019.

\bibitem{jar/LammichS19}
Peter Lammich and S.~Reza Sefidgar.
\newblock Formalizing network flow algorithms: {A} refinement approach in
  isabelle/hol.
\newblock {\em J. Autom. Reasoning}, 62(2):261--280, 2019.

\bibitem{LEDAsystem}
{LEDA} ({L}ibrary of {E}fficient {D}ata {T}ypes and {A}lgorithms).
\newblock \myurl{www.algorithmic-solutions.com}.

\bibitem{LEMON}
{LEMON} graph library.
\newblock COIN-OR project.

\bibitem{Leroy-JAR09}
Xavier Leroy.
\newblock A formally verified compiler back-end.
\newblock {\em Journal of Automated Reasoning}, 43:363--446, 2009.

\bibitem{McConnell2010}
R.M. McConnell, K.~Mehlhorn, S.~N\"{a}her, and P.~Schweitzer.
\newblock Certifying algorithms.
\newblock {\em Computer Science Review}, 5(2):119--161, 2011.

\bibitem{Mehlhorn-Naeher:LEDA}
K.~Mehlhorn and S.~N\"{a}her.
\newblock {LEDA}: A library of efficient data types and algorithms.
\newblock In {\em {MFCS}'89}, volume 379 of {\em {\rm Lecture Notes in Computer
  Science}}, pages 88--106, 1989.

\bibitem{Mehlhorn-Naeher4}
K.~Mehlhorn and S.~N\"aher.
\newblock \htmladdnormallink{The implementation of geometric
  algorithms}{http://www.mpi-sb.mpg.de/\~{}mehlhorn/ftp/ifip94.ps}.
\newblock In {\em Proceedings of the 13th {IFIP} World Computer Congress},
  volume~1, pages 223--231. Elsevier Science B.V. North-Holland, Amsterdam,
  1994.

\bibitem{Mehlhorn-Naeher:MFCS98}
K.~Mehlhorn and S.~N\"{a}her.
\newblock From algorithms to working programs: On the use of program checking
  in {LEDA}.
\newblock In {\em {MFCS}'98}, volume 1450 of {\em {\rm Lecture Notes in
  Computer Science}}, pages 84--93, 1998.

\bibitem{LEDAbook}
K.~Mehlhorn and S.~N\"aher.
\newblock {\em The {L}{E}{D}{A} {P}latform for {C}ombinatorial and {G}eometric
  {C}omputing}.
\newblock Cambridge University Press, 1999.

\bibitem{milner1972logic}
Robin Milner.
\newblock Logic for computable functions: description of a machine
  implementation.
\newblock Technical report, Stanford University, 1972.

\bibitem{jfp/MyreenO14}
Magnus~O. Myreen and Scott Owens.
\newblock Proof-producing translation of higher-order logic into pure and
  stateful {ML}.
\newblock {\em J. Funct. Program.}, 24(2-3):284--315, 2014.

\bibitem{nipkow2002isabelle}
Tobias Nipkow, Lawrence~C. Paulson, and Markus Wenzel.
\newblock {\em {Isabelle/HOL}---A Proof Assistant for Higher-Order Logic},
  volume 2283 of {\em {\rm Lecture Notes in Computer Science}}.
\newblock Springer, 2002.

\bibitem{noschinski:2011:graph-library}
Lars Noschinski.
\newblock A graph library for {I}sabelle.
\newblock {\em Mathematics in Computer Science}, 9:22--39, 2015.

\bibitem{Verification-CertComps-AutoCorres-Simpl}
Lars Noschinski, Christine Rizkallah, and Kurt Mehlhorn.
\newblock Verification of certifying computations through {A}uto{C}orres and
  {S}impl.
\newblock In {\em NASA Formal Methods}, volume 8430 of {\em LNCS}, pages
  46--61, 2014.

\bibitem{paulson1994isabelle}
Lawrence~C Paulson.
\newblock {\em Isabelle: A generic theorem prover}, volume 828.
\newblock Springer, 1994.

\bibitem{schirmer:2006:thesis}
Norbert Schirmer.
\newblock {\em Verification of Sequential Imperative Programs in
  {Isabelle/HOL}}.
\newblock PhD thesis, Technische Universit{\"a}t M{\"u}nchen, 2006.

\bibitem{Verisoft}
Verisoft.
\newblock \url{http://www.verisoft.de/}, 2007.

\bibitem{Boost}
Andrew~Lumsdaine von Jeremy G.~Siek, Lie-Quan~Lee.
\newblock {\em The {B}oost Graph Library: User Guide and Reference Manual}.
\newblock Addison-Wesley Professional, 2001.

\bibitem{Yap97}
C.-K. Yap.
\newblock Towards exact geometric computation.
\newblock {\em CGTA: Computational Geometry: Theory and Applications}, 7, 1997.

\bibitem{yap:crc:03}
C.-K. Yap.
\newblock Robust geometric computation.
\newblock In J.E. Goodman and J.~O'Rourke, editors, {\em Handbook of Discrete
  and Computational Geometry}, chapter~41. CRC Press LLC, Boca Raton, FL, 2nd
  edition, 2003.

\end{thebibliography}

\end{document}